\documentclass[aps,prl,reprint,superscriptaddress]{revtex4-2}

\usepackage[utf8]{inputenc}
\usepackage[T1]{fontenc}
\usepackage{lmodern}
\usepackage{graphicx}
\usepackage{enumerate}
\usepackage[explicit]{titlesec}
\usepackage{blindtext,color}
\usepackage{hyperref}
\usepackage{amsmath}
\usepackage{amssymb}
\usepackage{amsthm}
\usepackage{bm}
\usepackage {tensor}
\usepackage{ulem}
\usepackage{etoolbox}
\usepackage{enumitem}
\usepackage[english]{babel}
\usepackage{dsfont}
\usepackage{float}
\usepackage{slashed,braket}
\usepackage{stackengine}
\usepackage{mathrsfs}
\usepackage{verbatim}
\usepackage{array}
\usepackage{upgreek}
\usepackage{bbm}
\numberwithin{equation}{section}

\newcommand{\beq}{\begin{equation}}
\newcommand{\eeq}{\end{equation}}
\newcommand{\beqa}{\begin{eqnarray}}
\newcommand{\eeqa}{\end{eqnarray}}

\newcommand{\Tr}{\operatorname{Tr}}

\newcommand{\zz}[1]{{#1}}

\renewcommand{\theequation}{\arabic{equation}}

\begin{document}

\title{Emergent Hall viscosity in the integer quantum Hall phases of graphene-like systems}

\author{M. Selch}
\affiliation{Physics Department, Ariel University, Ariel 40700, Israel}

\author{M. A. Zubkov}
\affiliation{Physics Department, Ariel University, Ariel 40700, Israel}

\begin{abstract}
We explicitly distinguish Hall viscosity as defined relative to the strain field vs. relative to an emergent vielbein or metric field and discuss it for graphene-like systems. Aside from the gravitational or vielbein/metric related ``geometric'' Hall viscosity prevailing throughout the literature, a contribution proportional to the Hall conductivity, the ``electronic'' Hall viscosity, due to the emergent strain induced gauge field exists. We unify both contributions within the ``emergent'' Hall viscosity, determine it explicitly for graphene in the semimetal and Semenoff semiconducting phases, and in the latter case compare it to its non-relativistic limit. Under these circumstances two topological invariants enter the emergent Hall viscosity in the presence of translation and rotation symmetries. We employ Green function representations for the electronic Hall viscosity, quantized rather generically, and the geometric Hall viscosity, the latter being valid only under Lorentz (or Galilean) symmetry. We discuss experimental perspectives for extracting the emergent Hall viscosity.
\end{abstract}

\maketitle
\textit{Introduction.—}
Graphene as well as physical systems containing graphene have become reference materials for theoretical and experimental studies of a host of novel quantum effects as a consequence of their extensive past study and discovered exotic properties \cite{castroneto2009the,goerbig2011electronic}. Pristine graphene is well approximated by a semimetal with two physically distinct Fermi points located in its Brillouin zone. Integrated in a multilayer or 3D system or exposed to external fields and distinct chemical environments, graphene is known to become gapped allowing for distinct massive topological phases. There are several well-known microscopic mechanisms by which graphene may acquire a sizable band gap. The unit cell of graphene consists of two atomic sites as part of a honeycomb lattice structure defining two sublattices A and B. A staggered sublattice potential produces an asymmetry between the sublattices, breaks inversion symmetry and gives rise to a topologically trivial semiconducting phase, the Semenoff massive phase \cite{semenoff1984condensed}. It may be generated experimentally, e. g., by adsorption/chemical functionalization and/or placing graphene on a (non-magnetic) substrate.\par
The Hall viscosity is a non-dissipative viscous response to shear deformations in systems with broken time-reversal symmetry. It has emerged as a geometric response function, usually defined relative to a metric field \cite{abanov2014electromagnetic,gromov2014density,hoyos2012hall,cho2014geometry} which we term geometric Hall viscosity. It encodes topological information of the underlying quantum state complementary to the quantized Hall conductivity \cite{avron1995viscosity,avron1998odd}. Hall viscosity is intimately tied to the Berry curvature of the many-body wavefunctions in momentum space \cite{read2009non}. Its existence is particularly salient in gapped quantum fluids such as integer and fractional quantum Hall (FQH) states where it has been studied initially. It has been found that the Hall viscosity subtly influences electrochemical potentials locally on finite spatial distance scales, leading to measurable corrections in electrostatic profiles and current distributions \cite{hoyos2012hall,hoyos2014hall}. The Hall viscosity has been studied in the hydrodynamic regime \cite{alekseev2016negative, scaffidi2017hydrodynamic,delacretaz2017transport,pellegrino2017nonlocal} and specifically for pure graphene samples \cite{sherafati2016hall,sherafati2019hall}. The first experimental detection of Hall viscosity in graphene following this theoretical work has been reported in \cite{berdyugin2019measuring}. A complementary study in terms of an emergent strain gauge field parametrizing the effective coupling of acoustic phonons to electrons has been proposed in \cite{cortijo2015hall,heidari2019hall} which we will term the electronic Hall viscosity. We will show that both studies yield independent contributions necessary to arrive at a complete picture in the form of the emergent Hall viscosity. This distinct notion is crucial for systems comprising emergent gauge and vielbein or metric fields and directly relevant to experimental measurements.\par
A striking feature is the topology implied by incompressible quantum Hall states. On the one hand the geometric Hall viscosity to density ratio is found to be quantized in the presence of translational and rotational symmetry and proportional to the so-called Wen-Zee ``shift'' 
\cite{read2011hall}. The geometric Hall viscosity is furthermore intimately related to the Hall conductivity as established by Ward identities for Galilean invariant systems \cite{bradlyn2012kubo} and extended empirically to the Lorentz invariant case \cite{sherafati2016hall,sherafati2019hall}. On the other hand the electronic Hall viscosity is proportional to the system's Hall conductivity.\par 
This Letter establishes the notion of emergent Hall viscosity using effective field theory methods for systems topologically equivalent to semimetallic and Semenoff semiconducting graphene unifying earlier work.
We review essentials of the effective low energy field theory of (un)strained graphene together with the non-relativistic limit of the semiconducting phase. Linear response theory for the Hall conductivity and Hall viscosity within Wigner-Weyl calculus \cite{chernodub2017scale,zhang2020influence,zubkov2023effect,zhang2019hall,suleymanov2019wigner} is employed to derive explicit formulas in terms of Green functions and to evaluate and discuss the topology of the emergent Hall viscosity in integer quantum Hall phases. {We discuss experimental findings and explore future experimental perspectives regarding the emergent Hall viscosity and the possibility to determine the geometric and electronic Hall viscosities separately within electronic
transport}. 
Numerous details relevant throughout our discussion are provided in Supplementary Material \cite{supmat}.



\textit{Effective field theory of Dirac fermions in graphene-like systems.—}
Pristine graphene at half-filling is a semimetal exhibiting two distinct band touching points in its Brillouin zone which are gapped by a Semenoff mass enforcing a phase transition towards a semiconductor. At the level of graphene's crystal honeycomb lattice the Semenoff mass may be introduced as a staggered sublattice potential in the atomic unit cell of graphene which comprises two lattice sites. Sublattice symmetry breaking may be induced by placing graphene on a substrate. It is naturally realized in, e. g., pristine hexagonal boron nitride.
In the vicinity of its Fermi points the low energy field theory action of non-interacting graphene-like systems in Minkowski spacetime may be written in the Dirac form \cite{katsnelson2013euler}
\begin{align}
\nonumber S_M=&\int d^3xe\Psi^{\dagger}Q\Psi =\int d^3xe\Psi^{\dagger}(\omega -\mathcal{H})\Psi \\
=&\int d^3xe\bar{\Psi}(i\gamma^ae_a^{\mu}D_{\mu}-\gamma^0\mu +im\tau^3)\Psi 
\end{align}
with Dirac spinors $\Psi$, $\Psi^{\dagger}$ and $\bar{\Psi}$, vielbein field $e_a^{\mu}$, its inverse determant $e=\det^{-1}(e_a^{\mu})$ and covariant derivative $D_{\mu}=\partial_{\mu}+iA_{\mu}+i\delta A_{\mu}$. The Dirac theory features sublattice ($\sigma$), valley ($\tau$) and spin degrees of freedom. We may therefore introduce Pauli matrices subject to each of these spaces in general. Notice that the Dirac $\gamma$-matrices refer to sublattice space with the structure in valley and spin spaces being trivial for semimetallic graphene. This allows to include the valley and spin degrees of freedom simply by a degrees of freedom multiplicity factor $g_{sv}=4$. The Minkowski spacetime Dirac $\gamma$-matrices in sublattice space may be represented by $\gamma^0=i\sigma^3$, $\gamma^1=\sigma^2$, $\gamma^2=-\sigma^1$ which implies $\gamma^0\gamma^i=\sigma^i$,  $i=1,2$. Then as usual $\bar{\Psi}=\Psi^{\dagger}\gamma^0$. We assume metric signature $(-,+,+)$. The vierbein field is given by $e_a^0=\frac{1}{e}\delta_a^0$ as well as $e_a^i=\frac{v_0}{e}\delta_a^i$ with Fermi velocity $v_0$. The gauge field contributions refer to a vector potential $A_{\mu}$ inducing a constant, plane-orthogonal background magnetic field and an electromagnetic probe field $\delta A_{\mu}$, respectively. We included a chemical potential $\mu$ and a Semenoff mass term $m\tau^3$ as well. 
In Euclidean space the time coordinate $t$ is replaced by an imaginary time coordinate $\tau =it$. If we denote Euclidean quantities with a tilde-symbol, the temporal tensor components of action constituents are modified regarding temporal tensor components as a consequence of a Wick rotation. 
Taking proper account of such modifications the Euclidean Dirac $\gamma$-matrices are related to those in Minkowski spacetime via $\tilde{\gamma}^0=i\gamma^0$, $\tilde{\gamma}^i=\gamma^i$. The Euclidean space action $S_E$ is given by
\begin{align}
\nonumber -S_E=&\int d^3xe\Psi^{\dagger}\tilde{Q}\Psi =\int d^3xe\Psi^{\dagger}(i\omega -\tilde{\mathcal{H}})\Psi\\ 
=&\int d^3xe\bar{\Psi}(\tilde{\gamma}^a\tilde{e}_a^{\mu}\tilde{D}_{\mu}+\tilde{\gamma}^0\mu +m\tau^3)\Psi .\label{SED}
\end{align}
The conjugate spinor is defined by $\bar{\Psi}=\Psi^{\dagger}\tilde{\gamma}^0$.


Graphene subject to an external strain field is known to give rise to an emergent gauge field $\tau^3A_i^{s}$ and a fictitious ``gauge field'' $\Gamma_i^{s}$, $i=1,2$, as well as modifications to the emergent vielbein field (and therefore the emergent metric \cite{leyva2015generalizing,volovik2015emergent,zubkov2015emergent}). These fields follow from a symmetry analysis of a tight-binding model of graphene's underlying hexagonal honeycomb lattice. To be specific, given a three dimensional displacement field $(u_i(x),h(x))$ $(i=1,2)$ with in-plane components $(u_1(x),u_2(x))$ and out-of-plane component $h$ we define the strain fields
\begin{align}
\bar{u}_{ij}=\frac{1}{2}(\partial_iu_j+\partial_ju_i+\partial_ih\partial_jh),\,\,\,\,u_{ij}=\bar{u}_{ij}(h=0).
\label{fullstrainfield}
\end{align}
as well as the rotation field
\begin{align}
\omega_{ij}=\frac{1}{2}(-\partial_iu_j+\partial_ju_i).
\end{align}
The strain field $\bar{u}_{ij}$ is valid for general lattice distortions, while $u_{ij}$ captures in-plane strain only. In view of extracting the emergent Hall viscosity of graphene-like systems we may limit ourselves to in-plane strains from now on. In practice two coordinate systems are mainly used to describe strain effects in crystals. One such system is the accompanying or crystal coordinate system whose spatial coordinates $X^i$ $i=1,2$ are inert to the applied strain field $u_{ij}$. The latter enters explicitly via a coordinate transformation from the crystal coordinate system to the laboratory coordinate system with coordinates $x^i$ by the relation $x^i=u^i(X)+X^i$. While strain effects are derived and described more straightforwardly in the accompanying or crystal coordinate system, we will make use of the laboratory reference frame only, as it is this frame which is relevant to experiment. In order to obtain a consistent theory of strained graphene it is necessary to (i) expand the full theory in the vicinity of the Fermi points of the strained system which differs from that of the unstrained one already at leading order and (ii) represent the vielbein in the Dirac action in conformity with covariance principles prevalently employed in the context of high energy physics. In-plane strain induces the following modifications of the vielbein and metric fields 
\begin{align}
&e_a^i=v_0\Big(\delta_a^i\Big(1+\frac{\beta}{2}u_{kk}\Big)+(1-\beta )\delta_a^ju_{ij} {+\delta_a^j\omega_{ij}}\Big),\label{strainvielbein}\\
&g^{ij}=e_a^i\delta^{ab}e_b^j=v_0^2\Big(\delta_{ij}(1+\beta u_{kk})+2(1-\beta )u_{ij}\Big),\label{straininversemetric}\\
&g_{ij}=\frac{1}{v_0^2}\Big(\delta_{ij}(1-\beta u_{kk})+2(\beta -1)u_{ij}\Big)\label{strainmetric}
\end{align}
with $i,j,a=1,2$ and 
\begin{align}
e^\mu_0 = \Big(1+\frac{\beta}{2} u_{kk}\Big)\delta^\mu_0.\label{temporalvielbein}
\end{align}
It furthermore gives rise to the covariant derivative modification $D_i\to D_i+i\tau^3A_i^{s}+i\Gamma^s_i$ with emergent strain induced gauge fields \cite{juan2013gauge}
\begin{align}
A_i^{s}=\frac{\beta}{2a}K_{ijk}\epsilon_{kl} u_{jl},\,\,\,\,\Gamma^s_i=\frac{1}{2v_0}\partial_ju_{ij}.
\label{straingaugefields}
\end{align}
The lattice symmetry enters in terms of the coefficients $K_{ijk}$ with non-vanishing components given by $K_{111}=-K_{122}=-K_{212}=-K_{221}=1$. For a derivation see, e. g., \cite{volovik2015emergent} and references therein.
The parameter $a$ represents the fundamental lattice constant and $\beta$ is the Grüneisen parameter of graphene or more generally a graphene-like system. The latter characterizes the change in hopping strength $t$ when adjacent atoms experience a nonzero relative displacement. For isotropic in-plane strain one finds, e. g., $t\to t^{s}=t(1-\frac{\beta}{2}u_{ii})$. In the case of the vierbein, the strain induced modifications may be interpreted as a space and time dependent Fermi velocity $v_{ij}=v_0(\delta_{ij}(1+\frac{\beta}{2}u_{kk})+(1-\beta )u_{ij}+\omega_{ij})$. The fictitious gauge field $\Gamma_i^{s}$ is necessary to maintain hermiticity of the theory. The identity $i\sigma^j\sigma^3=\epsilon^{jk}\sigma^k$ may be used to show that $\Gamma_i^{s}$ takes the form of a connection for the $SO(2)$ group of local pseudospin rotations in sublattice space. The gauge field $A_i^{s}$ has opposite valley charges and therefore represents an axial gauge field in valley space.\par
Notice that the strain induced fictitious gauge field $\Gamma_i^{s}$ may be eliminated by employing a derivative symmetrized Dirac action with derivatives acting exclusively on the fermion fields. The above findings apply at linear order with $O(u_{ij}^2)$ correction terms neglected. We will make use of a gauge in spinor space such that $\omega_{ij}$ may be traded for inhomogeneous and higher order terms in strain fields. As we are interested in extracting strain effects in a homogeneous phase, we will from now on assume $u_{ij}=u_{ij}(t)$ such that $\Gamma_i^{s}=\omega_{ij}=0$. Furthermore our interest in the Hall viscosity allows us to neglect dilation effects. This leads to the further simplification $u_{ii}=0$ implying the tracelessness of the strain tensor as well as vielbein determinant $e=1$ within our approximations.
We refer to the relativistic fermions generally as Dirac fermions.

Next we will consider the non - relativistic limit of Eq. (\ref{SED}) as a standard non - relativistic approximation, which is valid for fermion energies much smaller than the effective Dirac mass. The expression for the Minkowski spacetime action of the resulting non-relativistic electrons of effective mass $m$ at chemical potential $\bar{\mu}=\mu -m$ receives the form
\begin{align}
\nonumber &S_M= \int d^3 x \sqrt{-g}\Psi^{\dagger}Q[\lambda] \Psi =\int d^3 x\sqrt{-g} \Psi^{\dagger}(\omega -\mathcal{H}) \Psi\\
&=\int d^3 x\sqrt{-g} \Psi^{\dagger}(iD_0+\bar{\mu} +\frac{1}{2m}g^{ij}D_iD_j+\frac{1}{2m}\tau^3eB)\Psi .\label{SMink}
\end{align}
The magnetic field induces a Zeeman term in valley space.\par
The Euclidean space action may be obtained by Wick rotation
\begin{align}
\nonumber &-S_E=\int d^3 x\sqrt{g} \Psi^{\dagger} \tilde{Q}[\lambda] \Psi =\int d^3 x\sqrt{g} \Psi^{\dagger}(i\omega -\tilde{\mathcal{H}}) \Psi\\
&=\int d^3 x\sqrt{g} \Psi^{\dagger}(-\tilde{D}_0+\bar{\mu} +\frac{1}{2m}g^{ij}D_iD_j+\frac{1}{2m}\tau^3eB)\Psi .\label{SEucl}
\end{align}
The non-relativistic fermions will be referred to as Pauli fermions \cite{volovik2013nambu,volovik2015scalar}.\par
We may add Coulomb interactions to the Dirac fermion approximation of graphene with the action of Eq. (\ref{SED}) and its non-relativistic limit comprising Pauli fermions with the action of Eq. (\ref{SEucl}) by replacing $\mu$/$\bar{\mu}$ by $\mu$/$\bar{\mu}$$-\lambda(x)$ and adding to the action the term 
\begin{align}
S[\lambda ]=&\frac{1}{2} \int dtd^3\bold{x} d^3\bold{x}^{\prime} \lambda(t,\bold{x}) V^{-1}(\bold{x},\bold{x}^{\prime })\lambda(t,\bold{x}^\prime).
\end{align}
Such a description is obtained naturally from a continuum limit derived from lattice regularized graphene in the presence of Coulomb interactions. The field $\lambda$ is the dynamical density fluctuation introduced by a Hubbard-Stratonovich transformation. It couples to fermions by a Yukawa term and mediates Coulomb interactions whose strength is determined by the Coulomb potential $V$ with inverse $V^{-1}=\frac{1}{e^2}\Delta^{(3)}$. This is nothing but the kinetic energy operator of the Hubbard-Stratonovich field. The operator $\Delta^{(3)}$ is the three dimensional Laplacian and $e$ is the electric charge.\par
We employ units such that $\hbar =c=e=1$ with $\hbar$ the reduced Planck constant, the velocity of light $c$ and the electric charge $e$. We will for notational convenience furthermore suppress the Fermi velocity $v_0$ for most of the discussion and omit tilde-symbols from Euclidean space quantities.\newline
\textit{Emergent Hall viscosity in linear response theory.—}
We consider field theory descriptions of fundamentally lattice regularized systems with itinerant electron-like particles which are spatially isotropic and homogeneous modulo the gauge fields giving rise to constant probe electric, strain rate and background magnetic fields.
We will now extract an expression for the emergent Hall viscosity in integer quantum phases and evaluate it explicitly for graphene-like systems in a generally Semenoff massive semiconducting phase and its non-relativistic limit. We begin with the definition of the strain stress tensor
\begin{align}
T^{s}_{ij}(x)=-\frac{\delta\log Z}{\delta u^{ij}(x)}=-\frac{1}{Z}\frac{\delta Z}{\delta u^{ij}(x)}.
\end{align}
with Euclidean partition function $Z$. We found that a homogeneous strain field gives rise to an emergent strain gauge field as well as modifications of the effective metric and vielbein fields (see Eqs. (\ref{strainvielbein}), (\ref{straininversemetric}), (\ref{strainmetric}) and (\ref{straingaugefields})). A derivation of the emergent Hall viscosity from the Euclidean partition function within linear response theory therefore involves a two-time application of the variational chain rule relating it to the Hall conductivity and the geometric Hall viscosity. The result for the emergent Hall viscosity is straightforward to obtain and reads \cite{supmat}
\begin{align}
\nonumber \eta^{em}_H=&\frac{1}{4}\epsilon^{ik}\delta^{jl}\frac{\partial \bar{T}_{ij}^s}{\partial (\partial_0u_{kl})}
=\frac{1}{2\pi}\Big[\Big(\frac{\beta}{2a}\Big)^2\mathcal{N}_{\sigma}+\\
&\frac{\beta (1-\beta )}{2a}(\mathcal{M}_1+\mathcal{M}_2)\sqrt{B}+(1-\beta )^2\mathcal{N}_{\eta}B\Big]
\label{topologicalemergenthallviscosity}
\end{align}
with out-of-plane background magnetic field $B$ and a bar $\bar{(\cdot )}$ indicating a configuration space average. The material Gr\"uneisen parameter $\beta$ is responsible for the response of hopping parameters to strain \cite{volovik2015emergent}. The calligraphic coefficients entering Eq. (\ref{topologicalemergenthallviscosity}) are expressed in terms of Weyl-symbols of operators. Of relevance are the fermion bilinear operator $\hat{Q}$, its inverse $\hat{G}$, which is the fermion propagator, and the covariant derivative. Weyl-symbols are functions on phase space which reduce to momentum space representations of operators in the homogeneous limit. A product of operators becomes a Moyal star product of Weyl-symbols.
For operators $\hat{A}$ and $\hat{B}$ we have $(\hat{A}\hat{B})_W=A_W\star B_W$ with 
\begin{align}
\star = {\rm exp} \Big(\frac{i}{2} \left( \overleftarrow{(\partial_x)}^{\mu}\overrightarrow{(\partial_p)}_{\mu}-\overleftarrow{(\partial_p)}_{\mu}\overrightarrow{(\partial_x)}^{\mu}\right) \Big).
\end{align}
For the fermion bilinear and its inverse this implies the Groenewold equation $Q_W\star G_W=1$ on phase space. We consider a homogeneous setup modulo the vector potential giving rise to a homogeneous background magnetic field. Therefore, the phase space dependence of operator Weyl symbols reduces to a dependence on the covariant derivative $i(D_W)_{\mu}=p_{\mu}-A_{\mu}(x)\equiv (\pi_W)_{\mu}$ or kinetic momentum. Note that the Moyal star product is associative but not commutative and is the phase space manifestation of operator ordering. In the presence of a background magnetic field not all momenta can be made good quantum numbers. Together with the reduced phase space dependence of Weyl symbols this is expressed by
\begin{align}
\star =
\exp\Big(-\frac{i}{2}B\epsilon_{ij}\overset{\leftarrow}{\partial}_{(\pi_W)_i}\overset{\rightarrow}{\partial}_{(\pi_W)_j}\Big).
\end{align}
More details on Wigner-Weyl calculus is provided in \cite{supmat}. In the general case the ``mixed'' coefficients $\mathcal{M}_i$ $i=1,2$ are naturally related to the physics of piezoelectricity \cite{vanderbilt2000berry,resta2007theory,droth2016piezoelectricity}. Piezoelectricity is not well defined in the integer quantum Hall regime due to its incompressibility \cite{bradlyn2012kubo}. We find that the coefficients $\mathcal{M}_i$ vanish identically in integer quantum Hall phases both for massless and massive fermions \cite{supmat}. 
The topological coefficients $\mathcal{N}_{\sigma}$ and $\mathcal{N}_{\eta}$ written in Weyl-symbol notation with $\omega =p_3$ read
\begin{align}
\nonumber \mathcal{N}_{\sigma}=&\frac{1}{24\pi^2\beta_T A}\int d^3xd^3p\epsilon_{\mu\nu\rho}\cdot\Tr\Big[\frac{\partial Q_W}{\partial p_{\mu}}\star \\
&G_W\star\frac{\partial Q_W}{\partial p_{\nu}}\star G_W\star \frac{\partial Q_W}{\partial p_{\rho}}\star G_W\Big],\label{topologicalhallcon}\\
\nonumber \mathcal{N}_{\eta}=&\frac{1}{16\pi^2\beta_T A}\int d^3xd^3p\frac{\epsilon^{ij}\delta^{lm}}{B}\Tr\Big[\frac{\partial Q_W}{\partial p_i}\star (D_W)_l\star \\
&G_W\star\frac{\partial Q_W}{\partial p_j}\star (D_W)_m\star G_W\star \frac{\partial Q_W}{\partial p_3}\star G_W\Big].\label{topologicalhallvis}
\end{align}
Similar expressions are obtained for $\mathcal{M}_1$ and $\mathcal{M}_2$, respectively. $\beta_T=\frac{1}{T}$ is the inverse temperature and we have assumed a square shaped sample with area $A=L^2$ and linear dimension $L$.
The coefficient $\mathcal{N}_{\sigma}$ represents the Hall conductivity Chern number such that the electron Hall conductivity is $\sigma_H=\frac{1}{2\pi}\mathcal{N}_{\sigma}$, while $\mathcal{N}_{\eta}$ represents the Hall viscosity topological invariant which is one fourth of the product of Wen-Zee shift $\mathcal{S}$ times the Hall conductivity Chern number {implying a geometric Hall viscosity $\eta^{geo}_H=\frac{1}{2\pi}\mathcal{N}_{\eta}B$.} In terms of the more common filling factor $\nu$, the Euler characteristic $\chi$, the electron-like charge carrier number $N$ and the number of flux quanta $N_{\Phi}=\frac{BA}{\Phi_0}$ with flux quantum $\Phi_0=2\pi$ the following relations are valid
\begin{align}
N=\nu N_{\Phi}+\mathcal{S}\frac{\chi}{2}\,\,\Rightarrow\,\,\mathcal{N}_{\sigma}=\nu ,\,\,\,\,\mathcal{N}_{\eta}=\frac{1}{4}\nu \mathcal{S}.
\end{align}
The first relation defines the Wen-Zee shift. On a spatial torus $\chi =0$. Note the relations $\nu_D=g_{sv}\Big(p-\frac{1}{2}\Big)$ for Dirac and $\nu_P=p-1$ for Pauli fermions for $p-1$ occupied Landau levels (above the unpaired Dirac Landau level in the former case).\newline
\textit{Topological robustness.—}
The Chern number $\mathcal{N}_{\sigma}$ given in Eq. (\ref{topologicalhallcon}) is known to be integer quantized quite generally at zero temperature. {The Chern number Green function representation in Eq. (\ref{topologicalhallcon}) is robust against arbitrary infinitesimal variations.} Its non-renormalization under perturbative Coulomb interactions in both relativistic and non-relativistic contexts has been discussed in \cite{qhemaik2,selch2026nonrenormalization} \zz{(for a general discussion of topological invariants see \cite{zubkov2018momentum,zubkov2012momentum,volovik2017standard,zubkov2012momentum,zubkov2017topology}). Non-perturbative effects give rise to the fractional quantum Hall effect \cite{qhemaik2}. The corresponding physics might have links to non-perturbative physics of QCD, where various topological defects dominate dynamics \cite{bakker1999central,abramchuk2018anatomy,bakker2005standard}.} The evaluation of the Chern number leads to
\begin{equation}
\mathcal{N}_{\sigma}=\sum_k\Theta (-\mathcal{E}_k)+\Delta\mathcal{N}_{\sigma}.
\label{halltopological}
\end{equation}
$\Theta$ is the Heaviside step function and $\mathcal{E}_k$ are energy eigenvalues relative to the chemical potential. The expression counts the number of occupied states. Note that the energy spectrum for Dirac fermions is not bounded from below. \zz{At the same time, the calculation based on emergent continuum field theory with Dirac fermions is not valid for sufficiently low energy levels. Hofstadter physics dominates in this regime, and the occupied levels contribute to the Hall conductivity with Chern numbers that may differ from $1$. It appears that there exist energy branches with large negative Chern numbers. They cancel the contributions of the branches below half filling with positive Chern numbers, so that effectively the Hall conductivity vanishes at half filling. }
{This implies the shift $\Delta\mathcal{N}_{\sigma}=1$ per spin degree of freedom and an effective counting of occupation relative to the unpaired Landau level.} For more details see \cite{Hatsugai} and references therein.
\par 
The coefficient $\mathcal{N}_{\eta}$ is only topological under more restrictive conditions. Rotational along with translational invariance imply the general functional forms \newline $Q_W=Q_W(\omega ,\gamma^ae_a^i(D_W)_i)$ for Dirac fermions and \newline $Q_W=Q_W\Big(\omega ,-\frac{(D_W)_ig^{ij}(D_W)_j}{2m}\Big)$ for Pauli fermions. Under these conditions $\mathcal{N}_{\eta}$ is known to be quantized. {The representation of $\mathcal{N}_{\eta}$ as given in Eq. (\ref{topologicalhallvis}) is valid only under full Lorentz symmetry (full Galilean symmetry in the non-relativistic case), though \cite{supmat}. Without Lorentz (or Galilean) boost symmetry only the leading coefficient in a filling factor expansion is able to faithfully represent the corresponding Hall viscosity contribution by the above Green function expression. A counterexample where Eq. (\ref{topologicalhallvis}) is inappropriate is the low energy effective theory of biased Bernal bilayer graphene which is described in terms of chiral fermions. This theory is neither Galilean nor Lorentz invariant \cite{selch2026valley}.} The explicit calculation of $\mathcal{N}_{\eta}$ leads to the expression
\begin{align}
\nonumber A\mathcal{N}_{\eta}=&\frac{iB}{4\pi} \sum_{k} \epsilon_{ij}\delta^{lm} \,\Big[ \langle k-2|{\hat \xi}_i|k-1\rangle \langle k-1| \hat{\xi}_l| k \rangle \cdot\\
\nonumber &\langle k | {\hat \xi}_j|k-1\rangle\langle k-1|\hat{\xi}_m | k-2\rangle \Big]\Theta(-\mathcal{E}_{k-2})\Theta(\mathcal{E}_k)\cdot\\
&\frac{(\mathcal{E}_{k-2}-\mathcal{E}_{k-1})\cdot (\mathcal{E}_k-\mathcal{E}_{k-1})}{(\mathcal{E}_k - \mathcal{E}_{k-2})^2}.
\label{hallviscosityintermediate}
\end{align}
The sum in $k$ is over discrete Landau levels and continuous intra-Landau level states as well as particle-hole conjugate partner states in the case of Dirac fermions. The kets $|k\rangle$ represent Dirac or Pauli fermion energy eigenstates and $-i\hat{D}_l=B\epsilon_{lm}\hat{\xi}_m$. The Heaviside step functions ensure that only the energy eigenfunctions and eigenvalues in a two Landau level neighborhood of the chemical potential need to be known for the determination of the Hall viscosity with a chemical potential in the gap modulo particle-hole conjugate states for Dirac fermions whose sum was suppressed in Eq. (\ref{hallviscosityintermediate}). Energy eigenvalues and eigenfunctions for Dirac and Pauli fermions may be determined generally in a straightforward way \cite{supmat,selch2026nonrenormalization}. 
{The robustness of the Green function representation of the Hall viscosity coefficient under infinitesimal variations respecting translational and rotational invariance to leading order in the filling factor has been shown in \cite{selch2026nonrenormalization}, while the full variational proof in the case of Lorentz and Galilean invariance is still missing.}

\textit{Explicit calculation of the emergent Hall viscosity.—}
{We employ a right-handed Cartesian coordinate system with unit vectors $(\hat{\bold{x}},\hat{\bold{y}},\hat{\bold{z}})$ such that $\bold{B}=B\hat{\bold{z}}$ and $B>0$. Our conventions are such that $\sigma_H=-\frac{1}{2}\epsilon^{ij}\sigma_{ij}$ as well as $\eta_H=-\frac{1}{4}\epsilon^{ik}\delta^{jl}\eta_{ijkl}$ with conductivity and viscosity tensors $\sigma_{ij}$ and $\eta_{ijkl}$ and $\epsilon^{xy}=-\epsilon^{yx}=1$.}\par
Both geometric Hall viscosity and Hall conductivity are odd under time-reversal symmetry and therefore under directional flip of the magnetic field $B\to -B$. They are furthermore odd under particle-hole conjugation {$p-\frac{1}{2}\to -p+\frac{1}{2}$ (with $p$ introduced below).} Semimetallic graphene at $\mu =0$ is particle-hole symmetric. Therefore its Hall conductivity as well as Hall viscosity vanish. Semenoff massive graphene has a particle-hole symmetric eigenvalue spectrum modulo the unpaired zero Landau level. Dirac fermions in graphene-like systems have opposite masses in the two valleys. The zero mode is shifted upward from zero in one valley but downward from zero in the other valley as soon as a Semenoff mass $m\tau^3$ is introduced. In the gapped phase we have a symmetry under particle-hole conjugation and simultaneous valley exchange. As we will sum over both valleys to obtain Hall conductivity and geometric Hall viscosity of graphene as a whole, this implies vanishing Hall conductivity and Hall viscosity within the gap $-|m|<\mu <|m|$. We may consider electron or hole doping relative to this gap and restrict ourselves to electron doping due to symmetry.
{We take the well-known anomalous contribution of the Hall conductivity of graphene $\Delta\mathcal{N}_{\sigma}=1$ per spin degree of freedom.}
The geometric Hall viscosity coefficient in Eq. (\ref{hallviscosityintermediate}) has been evaluated for Pauli fermions in \cite{selch2026nonrenormalization} and for Dirac fermions using symbolic python. 
The electron-doped case corresponds to $\mu >|m|$. For $|E_{p-1}|<\mu <E_{p}$ with $p\geq 1$ we obtain $\mathcal{N}_{\sigma}=(p+(p-1))$ per spin degree of freedom and $\mathcal{N}_{\eta}=\frac{1}{8}(p^2+(p-1)^2)$ per valley and spin degree of freedom.\par
The Pauli fermions have a valley dependent dispersion and therefore valley dependent quantized Landau level energy eigenvalues in a magnetic field with relative Landau level occupation of one for given chemical potential $\bar{\mu}$. If the first $p-1$ Landau levels above the unpaired Dirac Landau level are occupied then $\mathcal{N}_{\sigma}=p+(p-1)$ and $\mathcal{N}_{\eta}=\frac{1}{4}(p^2+(p-1)^2)$ per spin degree of freedom which coincides with the Dirac case.\par
Our final result for the emergent Hall viscosity including degeneracies {and doping sign $s_p=sgn\Big(p-\frac{1}{2}\Big)$} is
\begin{align}
\nonumber\eta_H^{em}=&\eta_H^{el}+\eta_H^{geo}=\frac{1}{2\pi}\Big(\frac{\beta}{2a}\Big)^2\mathcal{N}_{\sigma}+\frac{1}{2\pi}(1-\beta )^2\mathcal{N}_{\eta}B
\end{align}
\begin{align}
=\frac{\beta^2}{2\pi a^2}\Big(p-\frac{1}{2}\Big)+\frac{(1-\beta )^2}{4\pi}s_p\Big(p^2+(p-1)^2\Big)B.
\label{calculatedemhallvis}
\end{align}
In comparison to \cite{sherafati2016hall,sherafati2019hall} we find coincident Hall conductivities, but their geometric Hall viscosity is larger by a factor of four. {This follows directly from their definition of the stress tensor which we consider is twice as large as it should be. Proper normalization in our case is confirmed by our non-relativistic calculations.}\newline
\textit{Experimental measurement of electronic and geometric Hall viscosities.—}
{Three common techniques of graphene isolation and preparation are mechanical exfoliation, chemical vapor deposition (CVD) growth (each followed by transfer to a substrate like hexagonal boron nitride (hBN)) and the high-temperature sublimation technique of treating silicon carbide (SiC). Note that ideal massless Dirac cones as well as minimal (gauge field) coupling of acoustic phonons as presented are derived and expected to appear only in free standing high quality graphene. Both surface adsorption and substrate contact modify the phonon coupling constant ratio $\frac{\beta}{2a}$ as well as the Dirac mass $m$ and possibly introduce new effects. E. g., graphene on SiC and hBN is known to become gapped due to sublattice symmetry breaking. The Dirac mass as well as the background moir\'e potential may be tuned by misaligning graphene and hBN crystal axes relative to each other \cite{kim2018accurate}. Notice that we neglected the contribution of flexural phonons by restricting our discussion to in-plane strains. This may be justified by inversion symmetry and the implied quadratic dependence of the strain tensor on out-of-plane displacements (see Eq. (\ref{fullstrainfield})) of free standing graphene. Substrate induced inversion symmetry breaking allows for linear coupling of flexural phonons to electrons whose effect has recently been measured \cite{sadeghi2023tunable} and might be of general relevance. We consider sample preparation (of graphene or graphene-like Dirac materials) such that the Dirac characteristics with phonon gauge coupling are parametrically conserved (with tunable $\frac{\beta}{2a}$ and $m$) with negligible perturbations.}\par
We find that the geometric Hall viscosity is suppressed relative to the electronic Hall viscosity by a factor $\sim (1-\beta )^2\beta^{-2}eBa^2$ which was concluded as well in \cite{cortijo2015hall,heidari2019hall}. For pristine graphene $(1-\beta )^2\beta^{-2}\sim \frac{1}{4}$ and for magnetic fields of several Tesla giving rise to integer Hall plateaus $eBa^2\sim 10^{-3}$ in natural units. Taking $B=12$T for the $p=1$ integer quantum Hall magnetic field and $a\approx 0.25$nm for the lattice constant of graphene the suppression factor is $eBa^2\approx 1.1\cdot 10^{-3}$. The number of occupied Landau levels $p$ relative to the unpaired Landau level below a fixed chemical potential scales as $p\sim \frac{1}{B}$. This implies that $\sigma_H,\eta^{el},\eta^{geo}\sim \frac{1}{B}$. Naively extrapolating this finding beyond the achievable experimental resolution of Hall plateaus to magnetic fields of around $40$mT as used in \cite{berdyugin2019measuring} implies in accordance with Table I. in \cite{mechelen2019viscous} that in \cite{berdyugin2019measuring} Berdyugin et. al. measured the geometric Hall viscosity of electrons defined via the vielbein and metric fields, while the electronic Hall viscosity seems to be absent despite a finite Hall conductivity. Such an extrapolation is slightly compromised due to the finite temperature of the graphene sample used in the measurements of \cite{berdyugin2019measuring}, though. Proximity screening of long range Coulomb interactions by a metallic layer recently enabled the measurement of the Hall viscosity for electron hydrodynamic flow in graphene/hBN superlattices \cite{kim2025viscous} with conclusions in line with \cite{berdyugin2019measuring}. Further corroboration of this finding may come from surface magneto-optic Kerr effect spectroscopy measurements of the Hall viscosity following the line of reasoning as presented in \cite{mechelen2021optical}. 
The absence of the electronic Hall viscosity in the mentioned experiments confirms or will further solidify that it is not a transport coefficient relevant to the hydrodynamic flow of electrons as originally proposed in \cite{cortijo2015hall}.

\textit{Conclusions.—}In this Letter we defined the concept of emergent Hall viscosity as the quantity of direct relevance for condensed matter systems with emergent gauge, metric and vielbein fields in the effective field theory regime in the presence of strain. While the geometric Hall viscosity, usually discussed in the literature, arises from emergent geometry, the electronic Hall viscosity requires the presence of an emergent gauge field implying a finite Hall conductivity. We calculated the emergent Hall viscosity and discussed its topological stability and non-renormalization with respect to perturbative Coulomb interactions for integer quantum Hall phases of Dirac and Pauli fermions so long as the physical system under consideration is both homogeneous and isotropic. We considered both semimetallic graphene and semiconducting graphene in a topologically trivial Semenoff mass phase as an explicit example system, since the emergent Hall viscosity can be extracted experimentally for this system. {The dominant contribution to the emergent Hall viscosity is the electronic Hall viscosity in the effective field theory regime $eBa^2\ll \Phi_0$ with flux quantum $\Phi_0=2\pi$ which seems to be irrelevant for hydrodynamic electronic flow. Finally we claim the existence of a well-defined geometric valley Hall viscosity for graphene-like systems which may be calculated straightforwardly within our formalism \cite{selch2026valley}.}\par 


\textit{Author Contributions.}
M.S.: Conceptualization, Investigation, Formal Analysis,
Writing--Original Draft.
M.A.Z.: Supervision, Investigation, Formal Analysis, 
Writing--Review \& Editing.

\onecolumngrid

\appendix

\section*{Supplementary Material for ``Emergent Hall viscosity in the integer quantum Hall phases of graphene-like systems''}

\renewcommand{\theequation}{S\arabic{equation}}

\section{Appendix A: Spinorial gauge transformations}
\label{spinorgaugetransformations}

We provide a detailed account of convenient spinor redefinitions employed in the main text and their consequences on the sublattice induced Dirac structure of the two graphene valleys as well as the strain dependent vielbein within the kinetic piece comprising the strain induced antisymmetric form $\omega_{ij}=\frac{1}{2}(\partial_iu_j-\partial_ju_i)$. We will work in Minkowski spacetime and represent the spinor bilinear in the Dirac action via Pauli matrices using the Dirac $\gamma$-matrix representation $\gamma^0=i\sigma^3$, $\gamma^1=\sigma^2$ and $\gamma^2=-\sigma^1$.\par
Then the original valley structure of the fermion bilinear is $\Psi^{\dagger}\sigma^i\partial_i\Psi$ in one valley and $\Psi^{\dagger}(\sigma^{\ast})^i\partial_i\Psi$ in the other with equal Semenoff mass terms $m\Psi^{\dagger}\sigma^3\Psi$ in each valley \cite{castroneto2009the}. We may perform a valley polarized spinor rotation in one valley such that $\Psi\to \tilde{\Psi}=e^{-i\frac{\pi}{2}\sigma^1}\Psi$ which aligns the Dirac structures of the kinetic terms of both valleys at the expense of opposite masses in valley space. This follows directly from Pauli matrix commutation relations. We employ this form in the main text by writing $m\to\tau^3m$ with Pauli matrices $\tau$ acting in valley space.\par
After performing this step we employ the further rotation $\Psi\to\tilde{\Psi}=e^{-\frac{i}{2}\omega\sigma^3}$ common to both valleys with $\omega_{ij}=\omega\epsilon_{ij}$ \cite{juan2013gauge}. The mass terms remain invariant, while the kinetic terms will be modified by
\begin{align}
\nonumber \Psi^{\dagger}\sigma^i\partial_i\Psi\to\tilde{\Psi}^{\dagger}\sigma^i\partial_i\tilde{\Psi}=&\Psi^{\dagger}\sigma^i\partial_i\Psi-\omega \Psi^{\dagger}i\sigma^i\sigma^3\partial_i\Psi +O(\partial_i\omega) +O(\omega^2)\\
=&\Psi^{\dagger}\sigma^i\partial_i\Psi-\omega_{ij} \Psi^{\dagger}\partial_i\sigma^j\Psi +O(\partial_i\omega) +O(\omega^2)
\end{align}
We employed the Pauli matrix identity $i\sigma^i\sigma^3=\epsilon_{ij}\sigma^j$. Notice that the term proportional to $\omega$ enters with a sign opposite to that of the action with strain modulated vierbein $\gamma^ae_a^i\partial_i=\sigma^i\partial_i+O(u_{ij},\omega_{ij})$. We may therefore trade the term linear in $\omega$ (or $\omega_{ij}$) for terms of order $O(\partial_i\omega )$ and $O(\omega^2)$, respectively. The approximation linear in spatially homogeneous strain fields implies that the extra terms may be neglected. In conclusion $\omega$ may be gauged away within the employed approximation.

\section{Appendix B: Eigenvalue problem for massive Dirac fermions in a constant background magnetic field and their application to Semenoff mass gapped semimetals}
\label{eigenproblem}

\subsection{Appendix B1: Massive Dirac fermion problem}
\label{eigenproblem}

We review the essentials of the eigenvalue and eigenfunction problem of massive Dirac fermions in a constant external magnetic field in Landau gauge $A_x=0$, $A_y=Bx$. Introducing the dimensionless shifted coordinate $\xi =\sqrt{B}\Big(x-\frac{p_y}{B}\Big)$ as well as creation and annihilation operators $\hat{a}^{\dagger}=\frac{1}{\sqrt{2}}(\xi -\partial_{\xi})$ and $\hat{a}=\frac{1}{\sqrt{2}}(\xi +\partial_{\xi})$ such that $[\hat{a},\hat{a}^{\dagger}]=1$ we have to solve $\hat{Q}\Psi =0$ with
\begin{align}
-\hat{Q}=-\gamma^0(\gamma^ae_a^{\mu}D_{\mu}+\gamma^0\mu +m)=
\begin{pmatrix}
i\omega-\mu +m & -i\sqrt{2B}\hat{a} \\
i\sqrt{2B}\hat{a}^{\dagger} & i\omega -\mu -m
\end{pmatrix}.
\label{massivedirac}
\end{align}
The eigenspectrum may be obtained straightforwardly after analytic continuation $i\omega\to -E$. We will furthermore set $\mu =0$. Taking care of the single zero mode $E_0$ for $m=0$ we find (with Fermi velocity $v_0$ reinstated)
\begin{align}
E_{\pm n}=\pm \sqrt{m^2v_0^4+2Bv_0^2n}\,\,\,\,n\geq 1,\,\,\,\,E_0=-mv_0^2.\label{EDirac}
\end{align}
The corresponding eigenfunctions are given by
\begin{align}
\Psi_{\pm n}=(c_{\pm n}\phi_{n-1},d_{\pm n}\phi_n)^T\,\,\,\,n\geq 1,\,\,\,\,\Psi_0=(0,\phi_0)^T.
\end{align}
The functional form of the eigenfunctions may be derived from $\phi_0=N_0e^{ip_yy}e^{-\frac{1}{2}\xi^2}$ which fulfills $\hat{a}\phi_0=0$ and application of the ladder relations $\hat{a}\phi_n=\sqrt{n}\phi_{n-1}$ and $\hat{a}^{\dagger}\phi_n=\sqrt{n+1}\phi_{n+1}$, respectively. $N_0$ is a normalization constant such that the $\phi_n$ with $n\in\mathbb{N}_0$ are unit normalized. The coefficients $c_{\pm n}$ and $d_{\pm n}$ for $n\geq 1$ are functions of the magnetic field $B$ as well as the mass $m$ and given explicitly by
\begin{align}
c_{\pm n}=\sqrt{\frac{E_{\pm n}+m}{2E_{\pm n}}},\,\,\,\,d_{\pm n}=\mp i\sqrt{\frac{E_{\pm n}-m}{2E_{\pm n}}}.
\end{align}
The mass dependence makes the eigenfunctions with exception of the zero mode mass dependent. It is therefore remarkable that this mass dependence drops out of the Hall viscosity as calculated directly from the eigenfunctions for a chemical potential in the energy gap. This is unlike the case of the Hall conductivity where the eigenfunctions do not enter explicitly and only the number of occupied Landau level states relative to the zero mode matter. The latter also fully determine the Hall viscosity. Notice that the zero mode is shifted upward relative to zero for a positive mass and downward for a negative mass. For gapped graphene two Dirac cones would exist in the case $m=0$ at the $K$- and $K^{\prime}$-points of the Brillouin zone. A finite Semenoff mass then shifts the zero mode upwards in energy in one valley but downwards in the other.

\subsection{Appendix B2: Generalized massive Dirac fermion problem}

We consider now the most general rotationally covariant generalization of massive Dirac fermions in a homogeneous magnetic field. Most importantly, the Landau level quantization remains robust in this generalized scenario. This case is relevant for understanding the topological robustness of the Hall viscosity for Dirac fermions. 
In generalization of Eq. (\ref{massivedirac}) we assume the eigenvalue problem to be of the form (at vanishing chemical potential $\mu =0$)
\begin{align}
-\hat{Q}=-\gamma^0(\gamma^0D_0+C(\gamma^ae_a^{i}D_{i}\gamma^be_b^{j}D_{j})\gamma^ce_c^{k}D_{k}+M(\gamma^ce_c^{i}D_{i}\gamma^de_d^{j}D_{j}) +m)
\end{align}
with arbitrary functions $C$ and $M$.

Simplifying the functional argument for constant vielbein implies
\begin{align}
\nonumber \gamma^ae_a^iD_i\gamma^be_b^jD_j=&\frac{1}{2}\{\gamma^a,\gamma^b\}e_a^ie_b^jD_iD_j+\frac{1}{4}[\gamma^a,\gamma^b]e_a^ie_b^j[D_i,D_j]\\
=&\delta^{ij}D_iD_j-qB\sigma^3.
\end{align}
We assumed flat space conditions, kept the electric charge $q$ general and used Dirac $\gamma$-matrix algebra with our main text conventions for the explicit form of the $\gamma$-matrices. The eigenfunction ansatz
\begin{align}
\Psi_{\pm n}=(c^g_{\pm n}\phi_{n-1},d^g_{\pm n}\phi_n)^T\,\,\,\,n\geq 1,\,\,\,\,\Psi_0=(0,\phi_0)^T
\end{align}
implies that the functional argument of $C$ and $M$ may be replaced by $\delta^{ij}D_iD_j-qB\sigma^3\to -2qBn$ for all $n\in\mathbb{N}_0$ after acting on the eigenfunctions. Consequently we may write generically
\begin{align}
\nonumber -\hat{Q}\Psi_n=&-\gamma^0(\gamma^0D_0+C-(2qBn)\gamma^ae_a^{i}D_{i}+M-(2qBn) +m)\Psi_n\\
=&C-(2qBn)\sigma^3\Big(\frac{1}{C-(2qBn)}\gamma^0D_0+\gamma^ae_a^{i}D_{i}+\frac{M-(2qBn) +m}{C-(2qBn)}\Big)\Psi_n.
\end{align}
Let us abbreviate $C_n=C-(2qBn)$ and introduce the eigenfunction dependent mass terms
\begin{align}
m_n=\frac{M-(2qBn) +m}{C-(2qBn)},\,\,\,\,M_n=M-(2qBn) +m.
\end{align}
This implies
\begin{align}
-\hat{Q}\Psi_n=C_n\sigma^3\Big(\frac{1}{C_n}\gamma^0D_0+\gamma^ae_a^{i}D_{i}+m_n\Big)\Psi_n.
\end{align}
The eigenfunctions are then fully specified by
\begin{align}
c^g_{\pm n}=\sqrt{\frac{E^g_{\pm n}+M_n}{2E^g_{\pm n}}}=\sqrt{\frac{E_{\pm n}+m_n}{2E_{\pm n}}},\,\,\,\,d^g_{\pm n}=\mp i\sqrt{\frac{E^g_{\pm n}-M_n}{2E^g_{\pm n}}}=\mp i\sqrt{\frac{E_{\pm n}-m_n}{2E_{\pm n}}}.
\end{align}
The generalized energy eigenvalues are
\begin{align}
E^g_{\pm n}=\pm C_nE_{\pm n}=\pm C_n\sqrt{m_n^2v_0^4+2Bv_0^2n}\,\,\,\,n\geq 1,\,\,\,\,E^g_0=C_0E_0=-C_0m_0v_0^2.
\end{align}
As a result the generalized eigenstates may be obtained from the ordinary ones by a replacement of the Dirac mass $m$ by an eigenstate dependent mass $m_n$. The energy eigenvalues are rescaled by $C_n$. 

\section{Appendix C: The non-relativistic limit of massive valley electrons in graphene}

We derive subsequently the non-relativistic limit of massive valley electrons in graphene which may be interpreted as semimetals subjected to a Semenoff massive semiconducting phase. This limit is relevant for chemical potentials in the vicinity of the mass gap such that $0<\bar{\mu}=\mu -m\ll m$ for $m>0$ and $m\ll \mu -m=\bar{\mu}<0$ for $m<0$ as well as $B\ll m^2$. The quantity $\bar{\mu}$ may be interpreted as a non-relativistic, Pauli theory chemical potential relative to the Dirac theory chemical potential $\mu$.\par
We consider the formulation with the Dirac structure of the valleys related by $\sigma\,\leftrightarrow\,\sigma^{\ast}$ implying equal Semenoff masses in the two valleys. The eigenvalue problem for a spinor $\Psi =(\phi ,\chi)^T$ takes the form
\begin{align}
0=-\hat{Q}\Psi =(-E+\sigma^1\pi^1+\xi \sigma^2\pi^2+m\sigma^3)\Psi 
\label{eigenvalueproblem}
\end{align}
with $v_0=1$. $\xi =\pm 1$ marks the valley index and $\pi^i=iD_i$ denotes the kinetic momentum where $D_i=\partial_i-iqA_i$ is the covariant derivative in the presence of a background magnetic field $B$ represented by the vector potential $A_i$. The eigenvalue problem of Eq. (\ref{eigenvalueproblem}) has the component form
\begin{align}
(m-E)\phi +(\pi^1-i\xi\pi^2)\chi &=0\label{first}\\
(\pi^1+i\xi\pi^2)\phi -(m+E)\chi &=0\label{second}.
\end{align} 
In the non-relativistic limit the energy may be written as $-E=sm+\epsilon$ with $|\epsilon |\ll |m|$. The index $s$ distinguishes conduction ($s=+1$) and valence ($s=-1$) bands so that $q=-s$ in units of the elementary charge. In the main text we consider electron doping such that $q=-1$. In the case $s=+1$ ($s=-1$) $\phi$ ($\chi$) is small relative to $\chi$ ($\phi$) such that we may solve the system approximately up to $O(m^{-2})$ ($O(m^{-3})$ without an external electric field) corrections by plugging the solution of the approximated small component $\phi$ of Eq. (\ref{first}) with $-E+m\approx 2m$ into Eq. (\ref{second}) ($\chi$ of Eq. (\ref{second}) with $-E-m\approx -2m$ into Eq. (\ref{first})) which leads to
\begin{align}
&s=+1:\,\,\,\,\phi =-\frac{1}{2m}(\pi^1-i\xi \pi^2)\chi ,\,\,\,\,\Big(-\frac{1}{2m}(\pi^1+i\xi\pi^2)(\pi^1-i\xi\pi^2)+\epsilon \Big)\chi =0,\\
&s=-1:\,\,\,\,\chi =\frac{1}{2m}(\pi^1+i\xi\pi^2)\phi ,\,\,\,\,\Big(\epsilon +\frac{1}{2m}(\pi^1-i\xi\pi^2)(\pi^1+i\xi\pi^2)\Big)\phi =0.
\end{align}
We now make use of the commutation relation $[\pi^i,\pi^j]=\epsilon^{ij}iqB$ with $\epsilon^{12}=-\epsilon^{21}=1$. This implies the final result
\begin{align}
\epsilon =\frac{s}{2m}\boldsymbol{\pi}^2+\frac{1}{2m}\xi qB.
\label{nonrelativisticdispersion}
\end{align}
The first term on the right-hand side of Eq. (\ref{nonrelativisticdispersion}) is the standard quadratic non-relativistic dispersion supplemented by a second term representing a valley dependent Zeeman shift. This result may be obtained as well by moving $\xi$ in Eq. (\ref{eigenvalueproblem}) from $\sigma^2$ to $m$ which is equivalent by a unitary spinor rotation around the $x$-axis by $2\pi$. In the presence of a magnetic field the kinetic term is quenched into discrete Landau levels. We consider here electron doping which implies $\frac{1}{2m}\boldsymbol{\pi}^2\to \frac{-qB}{m}(n+\frac{1}{2})$ such that the energy spectrum acquires the form
\begin{align}
\epsilon_n =\frac{|qB|}{m}\Big(n+\frac{1}{2}+sgn(qB)\frac{\xi}{2}\Big)
\label{nonrelativisticeigenvalues}
\end{align}
The non-relativistic Landau levels of the two valleys are therefore shifted relative to each other by exactly one level which implies a relative Landau level occupation of one for a given chemical potential. The valley dependent Zeeman shift implies a shift of the fermion operator according to $\hat{Q}\to\hat{Q}-\frac{1}{2m}\xi qB$.

\section{Appendix D: The effective action and hydrodynamics for field theories with non-relativistic diffeomorphism invariance derived from the relativistic theory}
\label{nonrelfromrel}

We consider field theories which couple to an external spatial metric and an external Abelian gauge field and may be derived as the non-relativistic limit of a high energy theory. The relativistic extension is assumed to be both diffeomorphism invariant with the associated conserved energy momentum tensor $T_{\mu\nu}$ and gauge invariant with associated conserved electric current $j^{\mu}$. The high and low energy regimes are specified by $E\gg m$ and $E\ll m$, respectively, with $m$ being the mass of the relevant quasiparticles (we assume the existence of one species for simplicity).\par
The Minkowski spacetime partition function $\mathcal{Z}$ and the effective action $\mathcal{W}$, which depend on the external gauge fields, are related by
\begin{align}
\mathcal{Z}[A_{\mu},g_{\mu\nu}]=e^{i\mathcal{W}[A_{\mu},g_{\mu\nu}]}
\end{align}
with metric tensor $g_{\mu\nu}$ and gauge field $A_{\mu}$ and give rise to the conserved currents via the relations ($g=det(g_{\mu\nu})$)
\begin{align}
T_{\mu\nu}=\frac{2i}{\sqrt{g}}\frac{1}{\mathcal{Z}}\frac{\delta \mathcal{Z}}{\delta g^{\mu\nu}}=-\frac{2}{\sqrt{g}}\frac{\delta \mathcal{W}}{\delta g^{\mu\nu}},\,\,\,\,j^{\mu}=i\frac{1}{\mathcal{Z}}\frac{\delta \mathcal{Z}}{\delta A_{\mu}}=-\frac{\delta \mathcal{W}}{\delta A_{\mu}}.
\end{align}
Diffeomorphisms and gauge transformations leave the effective action invariant. The linear order variations of
\begin{align}
\mathcal{W}[A_{\mu}+\delta A_{\mu},g_{\mu\nu}+\delta g_{\mu\nu}]=\mathcal{W}[A_{\mu},g_{\mu\nu}]
\end{align}
imply the Minkowski spacetime ($g_{\mu\nu}=\eta_{\mu\nu}$) hydrodynamic equations
\begin{align}
\partial_{\mu}T^{\mu\nu}=0,\,\,\,\, \partial_{\mu}j^{\mu}=0\,\,\,\,(x^{\mu}=(ct,\bold{x})).
\end{align}
We may follow these relations from the relativistic theory at high energies to its low energy regime \cite{SonWingate2006}. We will assume a vanishing relativistic Abelian gauge field, as such a gauge field will be regenerated from the temporal components of the metric. In the non-relativistic limit we take $c\to \infty$ with $c$ being the (reinstated) velocity of light in vacuum and employ the (inverse) metric expansions
\begin{align}
g_{\mu\nu}=\begin{pmatrix}
-1-\frac{2A_0}{mc^2} & -\frac{A_i}{mc} \\
-\frac{A_i}{mc} & g_{ij}
\end{pmatrix},\,\,\,\,
g^{\mu\nu}=\begin{pmatrix}
-1+\frac{2A_0}{mc^2}+\frac{A_ig^{ij}A_j}{m^2c^2}+O(c^{-4}) & -\frac{g^{ij}A_j}{mc}+O(c^{-3}) \\
-\frac{g^{ij}A_j}{mc}+O(c^{-3}) & g^{ij}+O(c^{-2})
\end{pmatrix}.
\end{align}
The invariance under diffeomorphisms at high energies manifests itself in the form of the so-called non-relativistic diffeomorphism invariance or gauged Galilean invariance at low energies. Proceeding as in the relativistic case we obtain the hydrodynamic equations 
\begin{align}
\partial_0T^0_{\,\,\,k}+\partial_iT^i_{\,\,\,k}=0,\,\,\,\,\partial_0\rho +\partial_ij^i=0
\end{align}
where
\begin{align}
\nonumber &\rho =i\frac{1}{\mathcal{Z}}\frac{\delta \mathcal{Z}}{\delta A_0}=-\frac{\delta \mathcal{W}}{\delta A_0},\,\,\,\,j^i=i\frac{1}{\mathcal{Z}}\frac{\delta \mathcal{Z}}{\delta A_i}=-\frac{\delta \mathcal{W}}{\delta A_i},\,\,\,\,T^0_{\,\,\,k}=img_{ik}\frac{1}{\mathcal{Z}}\frac{\delta \mathcal{Z}}{\delta A_i}-iA_k\frac{1}{\mathcal{Z}}\frac{\delta \mathcal{Z}}{\delta A_0}=-mg_{ik}\frac{\delta \mathcal{W}}{\delta A_i}+A_k\frac{\delta \mathcal{W}}{\delta A_0},\\
&T^i_{\,\,\,k}=2ig_{kj}\frac{1}{\mathcal{Z}}\frac{\delta \mathcal{Z}}{\delta g_{ij}}-i\delta^i_k\log\mathcal{Z}+iA_k\frac{1}{\mathcal{Z}}\frac{\delta \mathcal{Z}}{\delta A_i}=-2g_{kj}\frac{\delta \mathcal{W}}{\delta g_{ij}}+\delta^i_k\mathcal{W}-A_k\frac{\delta \mathcal{W}}{\delta A_i}.
\end{align}
In order to extract the Hall viscosity from the spatial stress $T^i_{\,\,\,k}$ we only need to consider the traceless part. In the case of rotational invariance, which implies $T^0_{\,\,\,i}=j^i=0$, we may therefore assume with $g_{ij}=\delta_{ij}+\delta g_{ij}$ and $\delta^{ij}\delta g_{ij}=0$ to leading order that
\begin{align}
T^i_{\,\,\,k}=2ig_{kj}\frac{1}{\mathcal{Z}}\frac{\partial\mathcal{Z}}{\delta g_{ij}}\approx T^i_{\,\,\,k}=2i\delta_{kj}\frac{1}{\mathcal{Z}}\frac{\partial\mathcal{Z}}{\delta g_{ij}}\,\,\Leftrightarrow\,\,T_{ij}=2i\frac{1}{\mathcal{Z}}\frac{\partial\mathcal{Z}}{\delta g^{ij}}.
\end{align}

\section{Appendix E: The Hall conductivity and Hall viscosity in (2+1) dimensions}
\label{SectPreliminaries}

The Hall viscosity and Hall conductivity are response functions which emerge in the case of time-reversal symmetry breaking. Such a breaking may be induced internally by intrinsic (band topology) and extrinsic (scattering) effects as well as externally by the application of, e. g., a magnetic field. We assume the latter to be the only source of time-reversal symmetry breaking. Both transport phenomena are non-dissipative and may remain finite at zero temperature in impurity free samples. They stand out relative to their dissipative counterparts. We continue to present their precise definitions.\par

\subsection{Appendix E1: The Hall conductivity}

A transport current may be induced by the application of an external driving field. In the case of the Hall conductivity we have the electric current $J^k$ caused by the application of an electric field $E_l$ according to
\begin{align}
J^k=\sigma_L\delta^{kl}E_l+\sigma_H\epsilon^{kl}E_l
\label{ohmslaw}
\end{align}
with the electric field being the only source of isotropy breaking. Within the linear response approximation the (dissipative) longitudinal conductivity $\sigma_L$ and the Hall (or transverse) conductivity $\sigma_H$ are to be determined for an isotropic sample. Eq. (\ref{ohmslaw}) is nothing but Ohm's law. In a magnetic field applied orthogonally to the sample plane and strong compared to disorder broadening and large compared to temperature the Hall conductivity exhibits topological quantization. It may then be represented as $\sigma_H=\frac{1}{2\pi}\mathcal{N}_{\sigma}$ with $\mathcal{N}_{\sigma}$ being an integer (fraction) for the integer (fractional) quantum Hall effect. If the magnetic field is the only source of time-reversal symmetry breaking the Onsager relations imply $\sigma_L(B)=\sigma_L(-B)$ and $\sigma_H(B)=-\sigma_H(-B)$, respectively.

\subsection{Appendix E2: The geometric and emergent Hall viscosities}

The Hall viscosity on the other hand is associated with strain rate induced stresses. Small sample deformations may be parametrized by the displacement vector field $u_i$ with $i=1,2$ from which a strain $u_{ij}$ and a strain rate $\partial_0u_{ij}$ may be defined. The strain rate may be related to a fluid three velocity $v_i$ such that $v_i=\partial_0u_i\,\Rightarrow\,\partial_iv_j+\partial_jv_i=\partial_0u_{ij}$. Strain fields due to external forces are sometimes claimed to be captured exclusively by fluctuations of a(n emergent) metric field according to $g_{ij}\to g_{ij}+\delta g_{ij}$ with $\delta g_{ij}=\partial_iu_j+\partial_ju_i=2u_{ij}$. This geometric effect is to be supplemented by atomic bond strength renormalization effects for emergent metrics, though. Therefore we will in the following distinguish between emergent (em) and geometric (geo) viscoelastic effects. We will be interested only in the case of spatially homogeneous strains (and therefore also homogeneous strain rates) and metric deformations. These imply stresses which we parametrize more generally by 
\begin{align}
&T^{em}_{ij}=p\delta_{ij}-\lambda^{em}_{ijkl}u_{kl}-\eta^{em}_{ijkl}\partial_0u_{kl}+O((\partial_0)^2),\label{strdef}\\
&T^{geo}_{ij}=p\delta_{ij}-\lambda^{geo}_{ijkl}g_{kl}-\eta^{geo}_{ijkl}\partial_0g_{kl}+O((\partial_0)^2).\label{geodef}
\end{align}
The coefficients are the pressure $p$, the elastic modulus $\lambda^{em/geo}_{ijkl}=(\kappa^{em/geo})^{-1}\delta_{ij}\delta_{kl}$ with inverse compressibility $(\kappa^{em/geo})^{-1}=-V\partial_Vp$ and viscosity tensor $\eta^{em/geo}_{ijkl}$. The Hall viscosity forms part of the latter tensor as a scalar coefficient which arises as follows (we suppress superscripts momentarily). In a time reversal invariant theory the Onsager relations imply that $\eta_{ijkl}=\eta^S_{ijkl}=\eta^S_{klij}$ which, in the case of rotational invariance, allows for only two distinct transport coefficients, namely the shear viscosity $\eta$ and the bulk viscosity $\zeta$, according to
\begin{align}
\eta_{ijkl}=\eta (\delta_{ik}\delta_{jl}+\delta_{il}\delta_{jk})+(\zeta -\frac{2}{d}\eta )\delta_{ij}\delta_{kl}.
\end{align}
When time reversal invariance is broken, as occurs, e. g., for the Hall fluid characterized by the presence of a strong external magnetic field, the Onsager relations enable the appearance of an antisymmetric viscosity contribution
\begin{align}
\eta_{ijkl}=\eta^S_{ijkl}+\eta^A_{ijkl},\,\,\,\, \eta^A_{ijkl}=-\eta^A_{klij}.
\end{align}
In the case of rotational invariance the antisymmetric viscosity vanishes for $d>2$ with $d$ denoting the number of spatial dimensions. This case thus singles out $d=2$ for which an antisymmetric viscosity is allowed if parity is also broken. The so-called Hall viscosity coefficient may then finally be defined by
\begin{align}
\eta^A_{ijkl}=-\frac{\eta_H}{2}(\epsilon_{ik}\delta_{jl}+\epsilon_{jk}\delta_{il}+\epsilon_{il}\delta_{jk}+\epsilon_{jl}\delta_{ik})\,\,\Rightarrow\,\, \eta_H=-\frac{1}{4}\epsilon_{ik}\delta_{jl}\eta^A_{ijkl}.
\label{hallviscosityextraction}
\end{align}
If the magnetic field is the only source of time-reversal symmetry breaking the Onsager relations imply that $\eta (B)=\eta -(B)$, $\zeta (B)=\zeta -(B)$ and $\eta_H(B)=-\eta_H-(B)$, respectively.\par
In the presence of a vielbein field $e_a^i$ as relevant for (emergent) relativistic systems we may represent the geometric part and extract the Hall viscosity as follows
\begin{align}
(T^{geo})^a_{i}=p\delta_i^a-(\lambda^{geo})_{ij}^{ab}e_b^j-(\eta^{geo})_{ij}^{ab}\partial_0e_b^j+O((\partial_0)^2),
\label{geodef2}\,\,\,\,\eta_H=-\frac{1}{4}\epsilon^{ab}\delta^{ij}(\eta^{geo})^{ab}_{ij}.
\end{align}
An example of a fluid allowing for both a non-vanishing Hall conductivity and viscosity is the quantum Hall fluid. Notice a factor two difference relative to \cite{selch2026nonrenormalization} due to our convention, see Eq. (\ref{geodef}).\par
Remember that it suffices to assume the metric to have unit determinant $\det (g_{ij})=1$ and to be only slightly perturbed from the constant spatial Minkowski/Euclidean metric ($\eta_{ij}=\delta_{ij}$). We made the same assumption for the strain tensor. Therefore
\begin{align}
g_{ij}=\delta_{ij}+\delta g_{ij},\,\,\,\,\delta^{ij}\delta g_{ij}=0,\,\,\,\,\delta^{ij}u_{ij}=0.
\end{align}
This will be enough to extract the Hall viscosity. Notice that spatial indices may be raised and lowered trivially within the linear response regime.\par
We now come to the explicit relation between Hall viscosity defined relative to strain vs. metric perturbations in Semenoff massive graphene. These essentially arise due to the vielbein, metric and the strain gauge field which are all functions of the strain tensor \cite{leyva2015generalizing}. A variation of strain is therefore in a sense more fundamental and eventually defines viscoelastic responses in real samples. We must therefore ultimately extract the emergent Hall viscosity in order to make contact with experiment. Within a linear response theory treatment we will therefore make use of the chain rule for variations (using Eqs. (5) - (7) and (9) of the main text)
\begin{align}
&\frac{\delta}{\delta u_{ij}}=\frac{\delta A^s_k}{\delta u_{ij}}\frac{\delta}{\delta A^s_k}+\frac{\delta e_a^{k}}{\delta u_{ij}}\frac{\delta}{\delta e_a^k}=\frac{\beta}{2a}K_{kil}\epsilon_{lj}\frac{\delta}{\delta A^s_k}+\frac{(1-\beta )}{2}(\delta_a^i\delta^{jk}+\delta_a^j\delta^{ik})\frac{\delta}{\delta e_a^k}+\Big(\frac{\beta}{2}\delta^k_a\delta^{ij}\frac{\delta}{\delta e_a^k}\Big)\,\,\,\,\text{(relativistic case)}\label{chainrule1}\\
&\frac{\delta}{\delta u_{ij}}=\frac{\delta A^s_k}{\delta u_{ij}}\frac{\delta}{\delta A^s_k}+\frac{\delta g_{kl}}{\delta u_{ij}}\frac{\delta}{\delta g_{kl}}=\frac{\beta}{2a}K_{kil}\epsilon_{lj}\frac{\delta}{\delta A^s_k}+(\beta -1)(\delta^i_k\delta^j_l+\delta^i_l\delta^j_k)\frac{\delta}{\delta g_{kl}}-\Big(\beta \delta^{ij}\delta_{kl}\frac{\delta}{\delta g_{kl}}\Big) \,\,\,\, \text{(non-relativistic case)}.\label{chainrule2}
\end{align}
The final terms in Eqs. (\ref{chainrule1}) and (\ref{chainrule2}) in brackets arise from trace contributions irrelevant for the calculation of the Hall viscosity. We included them for completeness. Employing these relations we find that $\eta_H^{em}$ will be a sum of two terms each containing a polynomial in the Grüneisen parameter $\beta_T$ multiplied by one of the integers $\mathcal{N}_{\sigma}$ and $\mathcal{N}_{\eta}$. We relate $\eta_H^{em}$ explicitly to $\eta_H^{geo}$ and $\sigma_H$ in the main text. 

\section{Appendix F: The Weyl-symbol of an operator}
\label{WignerWeylcalculus}

We present here the basic definitions within the Wigner-Weyl calculus taken mostly from appendix B of \cite{qhemaik}. On the one hand it provides a bridge between the canoncial formulation of quantum mechanics in terms of operators and states on a Hilbert space and functions on phase space. It is an equivalent reformulation of the quantum physics on Hilbert spaces assuming Weyl-ordering of operators. A description by functions on phase space is familiar from classical physics and may be interpreted as getting deformed by quantum physics due to finite Planck constant $h$. In other words the quantum phase space formulation reduces to the classical one in the limit $h\to 0$. On the other hand the Wigner-Weyl transformation of an operator to a phase space function allows for an additional way of representing physics as compared to that in configuration or momentum space (which are related by Fourier transformation). Since in the main text we deal with effective continuous field theory of the quantum Hall effect systems, we consider here the Wigner-Weyl calculus adapted to continuous field theory. Notice, however, that this calculus remains valid as well almost without modifications for the true lattice models provided that the external field vary sufficiently slow in space. 
\par
The Wigner-Weyl calculus establishes a one-to-one correspondence between a quantum mechanical theory defined on a Hilbert space and its reformulation in terms of functions on phase space. An operator $\hat{\mathcal{O}}$ on Hilbert space, being a function of the position operator $\hat{x}$ and the momentum operator $\hat{p}$ in $D$ spacetime dimensions, is associated with a phase space function $O_W(p,x)$, being a function of space coordinate $x$ and momentum coordinate $p$, by the definition
\begin{align}
\hat{\mathcal{O}}=\int \frac{d^Dk}{(2\pi )^D}\frac{d^Dp}{(2\pi )^D}d^Dyd^DxO_W(x,p)e^{i(k(x-\hat{x})+y(p-\hat{p}))}.
\end{align}
A wave function on Hilbert space is represented in bra-ket notation by $|\Psi\rangle$ with configuration space representation $\Psi (x)=\langle x|\Psi\rangle$. We would like to relate the phase space function $O_W(x,p)$ to the configuration space representation of $\hat{\mathcal{O}}$ denoted by $O(x_1,x_2)=\langle x_1|\hat{\mathcal{O}}|x_2\rangle$ as well as to its momentum space representation denoted by $\tilde{O}(p_1,p_2)=\langle p_1|\hat{\mathcal{O}}|p_2\rangle$. The configuration and momentum space representations are related by
\begin{align}
O(x_1,x_2)=\int\frac{d^Dp_1}{(2\pi )^{\frac{D}{2}}}\frac{d^Dp_2}{(2\pi )^{\frac{D}{2}}}e^{ip_1x_1}\tilde{O}(p_1,p_2)e^{-ip_2x_2}.
\end{align}
In order to achieve this we calculate $\langle z|\hat{\mathcal{O}}|\Psi\rangle$. Applying the identities
\begin{align}
\langle z|e^{ik\hat{x}}|\Psi\rangle =e^{ikz}\langle z|\Psi\rangle =e^{ikz}\Psi (z),\,\,\,\, \langle z|e^{-iy\hat{p}}|\Psi\rangle =e^{-y\partial_z}\langle z|\Psi\rangle =e^{-y\partial_z}\Psi (z) =\Psi (z-y)
\end{align}
as well as the Baker-Campbell-Hausdorff relation
\begin{align}
e^{-i(k\hat{x}+y\hat{p})}=e^{-ik\hat{x}}e^{-iy\hat{p}}e^{i\frac{ky}{2}}
\end{align}
we obtain
\begin{align}
\nonumber \langle z|\hat{\mathcal{O}}|\Psi\rangle =&\langle z|\int \frac{d^Dk}{(2\pi )^D}\frac{d^Dp}{(2\pi )^D}d^Dyd^DxO_W(x,p)e^{i(k(x-\hat{x})+y(p-\hat{p}))}|\Psi\rangle\\
\nonumber =&\int \frac{d^Dk}{(2\pi )^D}\frac{d^Dp}{(2\pi )^D}d^Dyd^DxO_W(x,p)e^{i(kx+yp+\frac{ky}{2})}e^{-ikz}\Psi (z-y)\\
\nonumber =&\int \frac{d^Dp}{(2\pi )^D}d^Dyd^DxO_W(x,p)\delta^D(x+\frac{y}{2}-z)e^{iyp}\Psi (z-y)\\
=&\frac{1}{(2\pi )^D}\int d^Dpd^DyO_W(z-\frac{y}{2},p)e^{ipy}\Psi (z-y).
\end{align}
A comparison with the configuration space relation $\langle z|\hat{O}|\Psi\rangle =\int d^DxO(z,x)\Psi (x)$ for $x=z-y$ implies
\begin{align}
O(x,y)=\frac{1}{(2\pi )^D}\int d^DpO_W(\frac{x+y}{2},p)e^{ip(x-y)}.
\label{toderivestarproduct}
\end{align}
An inversion may be performed by a change of variables $R=\frac{x+y}{2}$ and $r=x-y$ followed by a Fourier transformation with respect to $r$. It reads
\begin{align}
O_W(R,p)=\int d^DyO(R+\frac{y}{2},R-\frac{y}{2})e^{-ipy}
\end{align}
The analogous relation for the momentum space representation may be obtained similarly and reads
\begin{align}
\tilde{O}(p,k)=\frac{1}{(2\pi )^D}\int d^DxO_W(x,\frac{p+k}{2})e^{i(k-p)x}
\label{momentumww}
\end{align}
with inverse
\begin{align}
O_W(R,p)=\int \frac{d^Dk}{(2\pi )^D}\tilde{O}(p+\frac{k}{2},p-\frac{k}{2})e^{ikR}.
\end{align}
The Weyl-symbol of the product $\hat{\mathcal{C}}=\hat{\mathcal{A}}\hat{\mathcal{B}}$ of two operators $\hat{\mathcal{A}}$ and $\hat{\mathcal{B}}$ on the Hilbert space may be represented as follows
\begin{align}
C_W(x,p)=A_W(x,p)\star B_W(x,p).
\label{weylsymbolproductofoperators}
\end{align}
The symbol $\star$ represents the Moyal star product
\begin{align}
\star = {\rm exp} \Big(\frac{i}{2} \left( \overleftarrow{(\partial_x)}^{\mu}\overrightarrow{(\partial_p)}_{\mu}-\overleftarrow{(\partial_p)}_{\mu}\overrightarrow{(\partial_x)}^{\mu}\right) \Big).
\end{align}
The form as written in Eq. (\ref{weylsymbolproductofoperators}) may be derived from Eqs. (\ref{toderivestarproduct}) and (\ref{momentumww}) and employing partial integration as follows
\begin{align}
\nonumber C_W(R,p)=&\int d^DyC(R+\frac{y}{2},R-\frac{y}{2})e^{-ipy}\\
\nonumber =&\int d^Dxd^DyA(R+\frac{y}{2},x)B(x,R-\frac{y}{2})e^{-ipy}\\
\nonumber =&\int d^Dxd^Dy\frac{d^Dq}{(2\pi )^D}\frac{d^Dk}{(2\pi )^D}A_W(\frac{R}{2}+\frac{y}{4}+\frac{x}{2},q)B_W(\frac{R}{2}-\frac{y}{4}+\frac{x}{2},k)e^{iq(R+\frac{y}{2}-x)}e^{ik(x-R+\frac{y}{2})}e^{-ipy}\\
\nonumber =&\int d^Dz_1d^Dz_2\frac{d^Dq}{(2\pi )^D}\frac{d^Dk}{(2\pi )^D}A_W(R+\frac{z_1}{2},q)B_W(R+\frac{z_2}{2},k)e^{-iqz_2}e^{ikz_1}e^{-ip(z_1-z_2)}\\
\nonumber =&\int d^Dz_1d^Dz_2\frac{d^Dq}{(2\pi )^D}\frac{d^Dk}{(2\pi )^D}A_W(R,q)e^{\frac{z_1}{2}\overset{\leftarrow}{\partial}_R}e^{\frac{z_2}{2}\overset{\rightarrow}{\partial}_R}B_W(R,k)e^{-iqz_2}e^{ikz_1}e^{-ip(z_1-z_2)}\\
\nonumber =&\int d^Dz_1d^Dz_2\frac{d^Dq}{(2\pi )^D}\frac{d^Dk}{(2\pi )^D}A_W(R,q)e^{ikz_1}e^{-\frac{i}{2}\overset{\leftarrow}{\partial}_k\overset{\leftarrow}{\partial}_R}e^{\frac{i}{2}\overset{\rightarrow}{\partial}_q\overset{\rightarrow}{\partial}_R}e^{-iqz_2}B_W(R,k)e^{-ip(z_1-z_2)}\\
\nonumber =&\int\frac{d^Dq}{(2\pi )^D}\frac{d^Dk}{(2\pi )^D}\Big[A_W(R,q)e^{\frac{i}{2}\overset{\rightarrow}{\partial}_k\overset{\leftarrow}{\partial}_R}e^{-\frac{i}{2}\overset{\leftarrow}{\partial}_q\overset{\rightarrow}{\partial}_R}B_W(R,k)\Big]\int d^Dz_1d^Dz_2e^{ikz_1}e^{-iqz_2}e^{-ip(z_1-z_2)}\\
\nonumber =&A_W(R,p)e^{\frac{i}{2}\overset{\rightarrow}{\partial}_p\overset{\leftarrow}{\partial}_R}e^{-\frac{i}{2}\overset{\leftarrow}{\partial}_p\overset{\rightarrow}{\partial}_R}B_W(R,p)\\
=&A_W(R,p)\star B_W(R,p).
\end{align}
We employed the substitutions $z_1=x+\frac{y}{2}-R,\,z_2=x-\frac{y}{2}-R$. If the operator $\hat{\mathcal{O}}$ is invertible, the identity
\begin{align}
O_W(x,p)\star O^{-1}_W(x,p)=1
\end{align}
on phase space is also known as the Groenewold equation of the Weyl-symbol of the operator $\hat{\mathcal{O}}$. Finally note that the Moyal star product is associative.\par
We will make use of the Groenewold equation to relate the Weyl-symbol of the propagator $G_W$ to the Weyl-symbol of the Dirac operator $Q_W$ according to
\begin{align}
Q_W(x,p)\star G_W(x,p)=1
\end{align}
on phase space. The Green function, or propagator, for time translation invariant theories or in thermodynamic equilibrium depends in general only on one (imaginary) time or frequency coordinate, while it depends on two spatial coordinates in spatial configuration, momentum or phase space. We define the Green function matrix element $G$ as the inverse of the Dirac operator matrix element $Q$ in configuration and momentum space as follows
\begin{align}
G(\bold{x},\bold{y},\omega )=\langle \bold{x}|(\omega -\hat{H})^{-1}|\bold{y}\rangle ,\,\,\,\,\tilde{G}(\bold{p},\bold{q},\omega )=\langle \bold{p}|(\omega -\hat{H})|\bold{q}\rangle .
\end{align}
We use bold letters in order to indicate coordinates only referring to spatial directions. $\hat{H}$ denotes the Hamilton operator of the physical theory under consideration. The corresponding Weyl-symbol of the propagator will be denoted by $G_W(\bold{x},\bold{p},\omega )$. The Feynman (or time ordered), Matsubara (or imaginary time ordered), retarded and advanced Green functions or propagators are defined as follows
\begin{align}
G^F_W(\bold{x},\bold{p},\omega )=G_W(\bold{x},\bold{p},\omega +i0^+sign(\omega )), \,\,\,\,G^M_W(\bold{x},\bold{p},\omega_n )=-iG_W(\bold{x},\bold{p},i\omega_n )\\
G^R_W(\bold{x},\bold{p},\omega )=G_W(\bold{x},\bold{p},\omega +i0^+),\,\,\,\, G^A_W(\bold{x},\bold{p},\omega )=G_W(\bold{x},\bold{p},\omega -i0^+)
\end{align}
with the superscripts distinguishing between the four types of propagators in the conventional way. The fermionic Matsubara frequencies are defined by $\omega_n=(2n+1)\pi T$ with $n\in\mathbb{Z}$. We denote a positive infinitesimal by $0^+$. We will usually make use of the Matsubara propagator and the associated Dirac operator and then leave the corresponding superscript implicit. The superscripts will be reinstated whenever necessary in the corresponding context. The Feynman prescription naturally corresponds to the path integral formulation in Minkowski spacetime, while the Matsubara prescription does so for the path integral formulation in Euclidean space. The retarded propagator relates to causal processes, while the advanced propagator relates to anti-causal processes. Within the Minkowski (Euclidean) path integral of field theory we find for the propagator the following expressions
\begin{align}
G^F(x,y)=-i\langle \psi (x)\bar{\psi}(y)\rangle\,\,\,\,(G^M(x,y)=\langle \psi (x)\bar{\psi}(y)\rangle )
\end{align}
in terms of the Grassmann-valued fermionic field variables $\Psi$ and $\overline{\Psi}$.

\section{Appendix G: Transport currents in Wigner-Weyl representation}
\label{WignerWeylCurrent}

We come now to the transport currents comprising Hall conductivity and Hall viscosity. The Euclidean space partition functions of massive Dirac electrons on the one hand and Pauli electrons on the other hand are given by
\begin{align}
Z = \int D\Psi D\bar{\Psi} D \lambda e^{S[\lambda] + \int d^3 x \bar{\Psi} \tilde{Q}[\lambda] \Psi}.
\label{ZEucl}
\end{align}
We will ultimately be interested in the respective field theory systems at low temperatures where Matsubara frequency sums may be effectively replaced by a frequency integration (in the deep quantum regime) and thus stick with the Euclidean path integral formulation. We will denote the Matsubara Green function (operator) by $G$ ($\hat{G}$), while the fermion (operator) (inverse propagator) is $Q$ ($\hat{Q}$) (we will omit the tilde from now on). Due to the presence of a constant magnetic (and possibly electric) field, our system is not homogeneous at the level of gauge fields. Momentum ceases to be a good quantum number. But since our system is still almost homogeneous, it is sensible to employ Wigner-Weyl calculus as a mean to expand around the strictly homogeneous theory. Notice that this formalism respects ordering descending from operator ordering via the Moyal star product. We present details of the employed Wigner-Weyl calculus in appendix \ref{WignerWeylcalculus}. Notice, though, that we use the so-called approximate version of the Wigner Weyl calculus, which is valid in the presence of weak inhomogeneities synonymous with the validity of an effective continuum field theory description.. Our definitions may be used without modifications for lattice systems when external fields vary slowly at distances of the order of the lattice spacing. Then the sum over the lattice points may everywhere be replaced by integrals. In practice this approximation works perfectly as long as the magnetic field strength is much smaller than around $10^5$ Tesla corresponding to $Ba^2\ll 1$ with magnetic field $B$ and lattice constant $a$ which is fulfilled in all laboratory accessible solid state systems up to artificial superlattice systems.\par
In order to extract the response functions of these systems, more specifically the Hall conductivity and the Hall viscosity, we consider variations of the Euclidean partition function in Eq. (\ref{ZEucl}) with respect to gauge and metric probe fields as follows.
The local electromagnetic current is given by the variation of the effective action with respect to the external electromagnetic gauge field. Under the assumption that the inverse propagator $Q$ is a function of the combination $p_{\mu}-A_{\mu}(x)$ only we find (note that the gauge field consists of the background magnetic field plus the probe field)
\begin{align}
J^k(x)=-\frac{\delta \log Z}{\delta A_k(x)}=-\frac{1}{Z}\frac{\delta Z}{\delta A_k(x)} =-\frac{1}{Z}\int D\Psi D\bar{\Psi} D\lambda e^{S[\lambda ] + \int d^3 x \bar{\Psi} Q[\lambda] \Psi} \bar{\Psi}(x) \frac{\partial}{\partial A_k}Q[\lambda] \Psi(x). 
\end{align}
Within operator and matrix element notation we obtain
\begin{align}
\nonumber J^k(x)=&\Tr[\hat{G}\frac{\delta}{\delta A_k(x)}\hat{Q}]=\int d^3y \Tr\Big[G(x,y)\frac{\partial Q(y,x)}{\partial A_k(x)}\Big]=-\int d^3k\frac{d^3p}{(2\pi)^{\frac{3}{2}}}\frac{d^3q}{(2\pi )^{\frac{3}{2}}}\Tr\Big[\tilde{G}(k,q)\frac{\partial}{\partial p_k}\tilde{Q}(q,p)\Big]e^{i(k-p)x}\\ =&-\int \frac{d^3p}{(2\pi )^3}\Tr\Big[G_W(x,p)\frac{\partial Q_W(x,p)}{\partial p_k}\Big].
\label{electriccurrent}
\end{align}
The subscript $W$ marks the Wigner-transform. 

The expectation value of the local stress $T_{ij}(x)$ of Pauli fermions is given by the variation of the partition function $Z$ with respect to the (inverse) metric probe field $g^{ij}(x)$ as
\begin{equation}
T_{ij}(x)=-2\frac{\delta \log Z}{\delta g^{ij}(x)}=-2\frac{1}{Z}\frac{\delta Z}{\delta g^{ij}(x)} =-\frac{2}{Z}\int D\Psi D\bar{\Psi} D\lambda e^{S[\lambda ] + \int d^3 x \bar{\Psi} Q[\lambda] \Psi} \bar{\Psi}(x) \frac{\partial}{\partial g^{ij}}Q[\lambda] \Psi(x). 
\end{equation}
This definition of local stress in the non-relativistic case is valid in rotationally invariant systems. We dropped here a potential metric contribution from $S[\lambda ]$ which is valid according to the results in \cite{selch2026nonrenormalization}. Within operator and Weyl-symbol notation this implies
\begin{equation}
T_{ij}(x)=2\Tr\Big[\hat{G}\frac{\delta}{\delta g^{ij}(x)}\hat{Q}\Big]=2\int \frac{d^3 p}{(2\pi)^3}\Tr\Big[G_W(x,p)\frac{\partial}{\partial g^{ij}} Q_W(x,p)\Big].
\end{equation}
Notice that $\hat{Q}$ and $\hat{G}$ are meant to contain a nonzero but homogeneous electric field in the case of the electric current and a nonzero and homogeneous stress rate $\partial_0g_{ij}$ in the case of the stress which we will extract by applying linear response theory. The response of the current to the electric field as well as the stress to the strain rate is reduced to the response of $\hat{G}$ to it in this approximation. $\hat{G}$ ($G_W$) is defined as a solution of the operator (Groenewold) equation
\begin{align}
\hat{G}\hat{Q}=\hat{Q}\hat{G}=\hat{1}\,\,\,\,(G_W \star Q_W=Q_W\star G_W=1).
\end{align}
Notice that $S[\lambda ]$ depends on the metric as well due to the kinetic operator. It has been shown in previous work that this term does not contribute to the stress \cite{selch2026nonrenormalization}.\par
The procedure for Dirac fermions follows the one presented for Pauli fermions very closely. The only difference is in the expression of the stress tensor which is defined by
\begin{equation}
T_i^a(x)=-\frac{\delta \log Z}{\delta e_a^i(x)}=-\frac{1}{Z}\frac{\delta Z}{\delta e_a^i(x)} =-\frac{1}{Z}\int D\Psi D\bar{\Psi} e^{S[\lambda ] + \int d^3 x \bar{\Psi}Q[\lambda ] \Psi} \bar{\Psi}(x) \frac{\partial}{\partial e_a^i}Q[\lambda] \Psi(x). 
\end{equation}
Within operator and Weyl-symbol notation we find
\begin{equation}
T_i^a(x)=\Tr\Big[\hat{G}\frac{\delta}{\delta e_a^i(x)}\hat{Q}\Big]=\int \frac{d^3 p}{(2\pi)^3}\Tr\Big[G_W(x,p)\frac{\partial}{\partial e_a^i} Q_W(x,p)\Big].
\end{equation}

\section{Appendix H: Linear response theory for Hall viscosity and Hall conductivity in terms of Green functions}
\label{LinearResponse}

The scheme to derive the Green function representation of response functions will be presented generically in the following but explained for definiteness for the geometric Hall viscosity of Pauli fermions. The replacements necessary to obtain the Hall conductivity for both Pauli and Dirac fermions or the Hall viscosity for Dirac fermions will be mentioned explicitly.

The expectation value of the (traceless part of the) stress tensor to a homogeneous strain rate within linear response theory averaged over the sample area $A$ at inverse temperature $\beta_T$ will be calculated subsequently. We intend more generally to vary the effective action $S_{eff}$ with respect to a probe field $X(x)$ in order to extract a configuration space averaged current $\bar{C}$ with the bar representing the average. In the case discussed here $\bar{C}=\bar{T}_{ij}$ and $X(x)=g^{ij}(x)$. We express the configuration space averaged current in terms of the Green function and its inverse within Wigner-Weyl representation. We performed this step explicitly in Section \ref{WignerWeylCurrent}. for both Pauli and Dirac fermions. These currents still contain a term linear in an external driving field $\partial_0Y(y)$ whose coefficient is the searched for response function. We will assume the driving field to be the time derivative of a probe field $Y(y)$, which is again the metric field here, as well as generally that $\frac{\delta^2\hat{Q}}{\delta X(x)\delta \partial_0Y(y)}=0$ for local probe fields $X(x)$ and $Y(y)$, respectively. Notice that the latter assumption is not generally valid (see, e. g., the fractional quantum Hall effect in \cite{selch2026nonrenormalization}) but sufficient for our purposes. This implies the validity of the following manipulations
\begin{align}
\nonumber \bar{T}_{ij}=&-\frac{2}{\beta_T A}\int d^3x\frac{\partial\log Z}{\partial g^{ij}}=\frac{2}{\beta_T A}\int d^3x\frac{1}{\beta_T}\sum_{\omega_n}\Tr\Big[\frac{\partial \hat{Q}^{\omega_n}}{\partial g^{ij}}\hat{G}^{\omega_n}\Big]\\
\nonumber =&-\frac{2}{\beta_T A}\int d^3x\frac{1}{\beta_T}\sum_{\omega_n}\Tr\Big[\frac{\partial \hat{Q}^{\omega_n}}{\partial g^{ij}}\hat{G}^{\omega_n}\frac{\partial \hat{Q}^{\omega_n}}{\partial (\partial_0g_{kl})}\hat{G}^{\omega_n}\Big]\partial_0g_{kl}\\
\overset{T\to 0}{=}&-\frac{2}{\beta_T A}\int d^3xi\int \frac{d\omega}{2\pi}\Tr\Big[\frac{\partial \hat{Q}^{\omega}}{\partial  g^{ij}}\hat{G}^{\omega}\frac{\partial \hat{Q}^{\omega}}{\partial (\partial_0g_{kl})}\hat{G}^{\omega}\Big]\partial_0g_{kl}.
\label{linearresponseresult1}
\end{align}
Note that we are interested in the limit of low temperatures where the Matsubara sum is well approximated by an angular frequency integral. The operators in the first line are assumed to be linear in the external driving field. Otherwise we will just keep this term and drop the persistent contribution. The operators in the second and third line are to be evaluated at vanishing probe fields. We complete our arguments now heuristically and formulate them afterwards more rigorously. The external driving field is not independent of the corresponding probe field but fulfills locally around a reference time $t_0$ the Taylor expansion relation $Y(y)=Y(t_0,\bold{y})+\partial_0Y(y)|_{t=t_0}(t-T_0)$ plus irrelevant terms. The probe field is assumed to be switched on smoothly at time $t_0$ such that $Y(t_0,\bold{y})=0$ but $\partial_0Y(y)|_{t=t_0}\neq 0$. Therefore we have locally in time $\frac{\delta}{\delta \partial_0Y(y)}=\frac{\delta}{\delta Y(y)}(t-t_0)$. We then interpret time as an operator in frequency space in Euclidean time such that $(t-t_0)=-i(\tau -\tau_0)=-i\frac{\partial}{\partial\omega}$.  Together with $\frac{\partial\delta\hat{Q}}{\partial\omega\delta Y(y)}=0$ we find
\begin{align}
\nonumber \bar{T}_{ij}=&\frac{2}{\beta_T A}\int d^3x\int \frac{d\omega}{2\pi}\Tr\Big[\frac{\partial \hat{Q}^{\omega}}{\partial  g^{ij}}\hat{G}^{\omega}\frac{\partial \hat{Q}^{\omega}}{\partial g_{kl}}\frac{\partial}{\partial \omega}\hat{G}^{\omega}\Big]\partial_0g_{kl}\\
\nonumber =&-\frac{2}{\beta_T A}\int d^3x\int \frac{d\omega}{2\pi}\Tr\Big[\frac{\partial \hat{Q}^{\omega}}{\partial  g^{ij}}\hat{G}^{\omega}\frac{\partial \hat{Q}^{\omega}}{\partial g_{kl}}\hat{G}^{\omega}\frac{\partial \hat{Q}^{\omega}}{\partial \omega}\hat{G}^{\omega}\Big]\partial_0g_{kl}\\
\nonumber =&-\frac{2}{\beta_T A}\int d^3x\int \frac{d^3p}{(2\pi)^3}\Tr\Big[\frac{\partial Q_W(x,p)}{\partial g^{ij}}\star G_W(x,p)\star\\
&\frac{\partial Q_W(x,p)}{\partial g^{kl}}\star G_W(x,p)\star\frac{\partial Q_W(x,p)}{\partial \omega}\star G_W(x,p)\Big]\partial_0g_{kl}\label{linearresponseresult2}\\
\nonumber \equiv &-\eta^{geo}_{ijkl}\partial_0g_{kl}.
\end{align}
Alternatively, due to the symmetry properties of the Hall viscosity, we may replace $\hat{G}^{\omega}\frac{\partial \hat{Q^{\omega}}}{\partial (\partial_0g_{kl})}\hat{G}^{\omega}$ by the  antisymmetrized product $\frac{i}{2}\Bigl(\hat{G}^{\omega}\frac{\partial \hat{Q}^{\omega}}{\partial g_{kl}} \partial_\omega\hat{G}^{\omega} - \partial_\omega \hat{G}^{\omega} \frac{\partial \hat{Q}^{\omega}}{\partial g_{kl}}\hat{G}^{\omega}\Bigr)$
\begin{align}
	\nonumber \bar{T}_{ij}=&-\frac{1}{\beta_T A}\int d^3x\int \frac{d^3p}{(2\pi)^3}\Tr\Big[\frac{\partial Q_W(x,p)}{\partial  g^{ij}}\star G_W(x,p)\star\\
	&\frac{\partial Q_W(x,p)}{\partial g^{kl}}\star G_W(x,p)\star\frac{\partial Q_W(x,p)}{\partial \omega}\star G_W(x,p)\Big]\partial_0g_{kl} - (\partial /\partial g_{kl}\leftrightarrow \partial/\partial \omega)
\end{align}
More rigorously we may consider an iterative solution of the Groenewold equation with time dependent fermion bilinear operator $Q_W^F=Q_W(g_{kl}(t))$ and metric field $g_{kl}(t)=\delta_{kl}+\delta g_{kl}(t)$ with further dependencies suppressed. There are two expansion schemes: linear response (L) in $\delta g_{kl}(t)$ and the gradient expansion (G) in powers of time derivatives. Note that the viscosity tensor is the piece multiplied by the first time derivative of the metric (or strain or vielbein) field only.  Set $Q_W^{L,0}=Q_W(\delta_{kl})$ and employ a linear response expansion/gradient expansion in time for the propagator Weyl-symbol $G_W(t)$, again with suppression of further arguments, such that $G_W(t)=\sum_{n=0}^{\infty}G_W^{L/G,n}$. At lowest order we set $G_W^{L,0}\star Q_W^{L,0}=1$ and $G_W^{G,0}\star Q_W^F=1$, respectively. In the time dependent case we have the star product generalization $\star\to\star \circ$ with
\begin{align}
G_W(t)\star \circ Q_W(g_{kl}(t)) = 1 , \quad \circ = e^{\frac{i}{2}\Big((\overleftarrow{\partial}_t)(\overrightarrow{\partial}_{\omega})-(\overleftarrow{\partial}_{\omega})(\overrightarrow{\partial}_t)\Big)}.
\end{align}
As we are interested in the static limit of the viscous Hall response in linear response theory, we obtain at first order
\begin{align}
1=\Big[G_W^{L/G,1}\star Q_W^{L,0/F}+G_W^{L/G,0}\star \Big(1+\frac{i}{2}\Big((\overleftarrow{\partial}_t)(\overrightarrow{\partial}_{\omega})-(\overleftarrow{\partial}_{\omega})(\overrightarrow{\partial}_t)\Big)\Big)Q_W(g_{kl}(t))\Big]\Big\rvert_{g_{kl}(t)=\delta_{kl}}.
\end{align}
Multiplication from the right by $\star G_W^{L/G,0}$ together with the time dependence of $G_W^{L,0}$ through the argument $g_{kl}(t)$ and both $G_W^{L,0}\star Q_W^{L,0}=1$ and $G_W^{G,0}\star Q_W^F=1$ leads to
\begin{align}
G_W^{L,1}=\frac{i}{2}\frac{\partial G_W^{L,0}}{\partial \omega}\star \frac{\partial Q_W^{L,0}}{\partial g_{kl}}\partial_0g_{kl}\star G_W^{L,0}=-\frac{i}{2}G_W^{L,0}\star\frac{\partial Q_W^{L,0}}{\partial \omega}\star G_W^{L,0}\star \frac{\partial Q_W^{L,0}}{\partial g_{kl}}\star G_W^{L,0}\partial_0g_{kl}.
\end{align}
on the one hand. Exploiting the antisymmetry $\eta^{geo, A}_{ijkl}=-\eta^{geo, A}_{klij}$ of the Hall viscous contribution a zero temperature limit of the first equality in Eq. (\ref{linearresponseresult1}) expressed in Weyl-symbols with $G_W^{L,1}$ for the linear approximation of the propagator leads directly to the third or penultimate equality of Eq. (\ref{linearresponseresult2}). On the other hand we find 
\begin{align}
&G_W^{G,1}\rvert_{g_{kl}(t)=\delta_{kl}}=\frac{i}{2}\frac{\partial G_W^{L,0}}{\partial \omega}\star \frac{\partial Q_W^{L,0}}{\partial g_{kl}}\partial_0g_{kl}\star G_W^{L,0} - \frac{i}{2}\frac{\partial G_W^{L,0}}{\partial g_{kl}}\star \frac{\partial Q_W^{L,0}}{\partial \omega}\partial_0g_{kl}\star G_W^{L,0} \\ &=-\frac{i}{2}G_W^{L,0}\star\frac{\partial Q_W^{L,0}}{\partial \omega}\star G_W^{L,0}\star \frac{\partial Q_W^{L,0}}{\partial g_{kl}}\star G_W^{L,0}\partial_0g_{kl} - (\partial \omega \leftrightarrow \partial g_{kl}).
\end{align}
Notice that the gradient expansion in time evaluated at $g_{kl}(t)=\delta_{kl}$ resembles our result for linear response. For the Hall viscosity we may replace $G^{L,1}\to \frac{1}{2}G^{G,1}$ due to symmetry which will appear as well in the following sections. More general formulae for the gradient expansion in time may be obtained using the Keldysh technique, when the above expressions comprise $2\times 2$ Keldysh Green functions. In thermal equilibrium they are reduced to those containing Matsubara Green functions. At zero temperature we therefore arrive at 
\begin{align}
	\nonumber \bar{T}_{ij}=&-\frac{1}{\beta_T A}\int d^3x\int \frac{d^3p}{(2\pi)^3}\Tr\Big[\frac{\partial Q_W(x,p)}{\partial  g^{ij}}\star G_W(x,p)\star\\
	&\frac{\partial Q_W(x,p)}{\partial g^{kl}}\star G_W(x,p)\star\frac{\partial Q_W(x,p)}{\partial \omega}\star G_W(x,p)\Big]\partial_0g_{kl} - (\partial/\partial g_{kl}\leftrightarrow \partial/\partial \omega)
\end{align}
For the antisymmetric viscosity this leads to the further expression.
\begin{align}
\eta^{geo}_H=&\frac{1}{4}\epsilon^{ik}\delta^{jl}\frac{\delta \bar{T}_{ij}}{\delta \partial_0g^{kl}}\\
=&-\frac{1}{2\beta_T A}\int d^3x\int \frac{d^3p}{(2\pi )^3}\epsilon^{ik}\delta^{jl}\Tr\Big[\frac{\delta Q_W(x,p)}{\delta g^{ij}}\star G_W(x,p)\star\\
&\frac{\delta Q_W(x,p)}{\delta g^{kl}}\star G_W(x,p)\star \frac{\partial Q_W(x,p)}{\partial \omega}\star G_W(x,p)\Big].
\end{align}
Following the same steps for the Hall conductivity of both Pauli and Dirac electrons and the Hall viscosity of Dirac electrons yields
\begin{align}
&\bar{J}^k=-\frac{1}{\beta_T A}\int d^3x\frac{1}{Z}\frac{\delta\log Z}{\delta A_k(x)}=\sigma_L\delta^{kl}E_l+\sigma_H\epsilon^{kl}E_l,\\
&\bar{T}_i^a=-\frac{1}{\beta_T A}\int d^3x\frac{1}{Z}\frac{\delta\log Z}{\delta e_a^i(x)}=-(\eta^{geo})^{ab}_{ij}\partial_0e_b^j
\end{align}
with
\begin{align}
\nonumber \sigma_H=&{\frac{1}{2\beta_T A}}\int d^3x \int \frac{d^3p}{(2\pi )^3}\epsilon_{ij}\Tr\Big[ \frac{\delta
Q_W(x,p)}{\delta A_i}\star G_W(x,p)\nonumber \\
&\star \frac{\delta Q_W(x,p)}{\delta A_j}\star G_W(x,p)\star \frac{\partial Q_W(x,p)}{\partial \omega}\star G_W(x,p)\Big]
\end{align}
and
\begin{align}
\nonumber \eta_H^{geo}=&-\frac{1}{4\beta_T A}\epsilon_{ab}\delta^{ij}\int d^3x\int \frac{d^3p}{(2\pi)^3}\Tr\Big[\frac{\delta Q_W(x,p)}{\delta  e_a^i}\star G_W(x,p)\star\\
&\frac{\delta Q_W(x,p)}{\delta e_b^j}\star G_W(x,p)\star\frac{\partial Q_W(x,p)}{\partial \omega}\star G_W(x,p)\Big].
\end{align}

\subsection{Appendix I: Topological invariance of viscous transport for systems of Pauli and Dirac fermions}
\label{TopologicalInvariants}

We now specify the fermion operators of our interest in order to manipulate their derivatives further and employing our results from appendix \ref{LinearResponse}. We start again with Pauli fermions. First notice that we will assume the fermion operator $\hat{Q}$ and its corresponding Weyl symbol to depend only on the gauge invariant combination $i(D_W)_{\mu}=p_{\mu}-A_{\mu}(x)\equiv (\pi_W)_{\mu}$. In other words $Q_W=Q_W(D_W)$ where $D_W$ is the Weyl symbol of the covariant derivative. Specifically for considerations of the Hall viscosity we assume that the system is homogeneous modulo the vector potential giving rise to a constant external background magnetic field as well as rotationally invariant. This implies that $Q_W=Q_W((D_W)_ig^{ij}(D_W)_j)$. These restrictions may be lifted for considerations of the Hall conductivity which is quantized under more generic conditions such as weak disorder and inhomogeneous magnetic fields. The fermion operator derivatives then fulfill the relations
\begin{align}
&\frac{\partial Q_W}{\partial g^{ij}}=\frac{1}{2m}(D_W)_{(i}\star (D_W)_{j)},\,\,\,\,-\frac{\partial Q_W}{\partial A_i}=\frac{\partial Q_W}{\partial p_i}=-\frac{i}{m}g^{ij}(D_W)_j,\,\,\,\,(D_W)_{\mu}=-ip_{\mu}+iA_{\mu}(x)\\
&\Rightarrow \frac{\partial Q_W}{\partial g^{ij}}=\frac{i}{2}\frac{\partial Q_W}{\partial p_{(i}}\star (D_W)_{j)}\Big\rvert_{g^{ij}=\delta^{ij}}=\frac{i}{2}(D_W)_{(i}\star\frac{\partial Q_W}{\partial p_{j)}}\Big\rvert_{g^{ij}=\delta^{ij}}
\label{metrictomomentum}
\end{align}
modulo commutators of Weyl symbols of operators with respect to the Moyal star product. Exact fulfillment requires the presence of Galilean boost symmetry implying the assumption of full Galilean symmetry. This has been shown explicitly in \cite{selch2026nonrenormalization}. The above derived expressions yield the following for the stress averaged over the area of the system in the linear response approximation (which allow us to set $g_{ij}=\delta_{ij}$ in the subsequent expressions)
\begin{align}
\nonumber \bar{T}_{ij}=&\frac{1}{4\beta_T A}\int d^3x\int \frac{d^3p}{(2\pi)^3}\Tr\Big[\frac{\partial Q_W(x,p)}{\partial p_{(i}}\star (D_W)_{j)}\star G_W(x,p)\star\\
&\frac{\partial Q_W(x,p)}{\partial p_k}\star (D_W)_{l}\star G_W(x,p)\star\frac{\partial Q_W(x,p)}{\partial \omega}\star G_W(x,p)\Big]\partial_0g_{kl} -(\partial /\partial g_{kl}\leftrightarrow \partial/\partial \omega).
\label{hallviscositynoninteracting}
\end{align}
The geometric Hall viscosity may then be determined with the aid of Eqs. (\ref{geodef}) and (\ref{hallviscosityextraction}), respectively. We obtain
\begin{align}
\eta^{geo}_H=\frac{1}{4}\epsilon^{ik}\delta^{jl}\frac{\delta \bar{T}_{ij}}{\delta \partial_0g^{kl}}=\frac{1}{2\pi}\mathcal{N}_{\eta}B
\end{align}
with external background magnetic field $B$ and ($p_3=\omega$)
\begin{align}
\nonumber \mathcal{N}_{\eta}=&\frac{1}{32\pi^2\beta_T A}\int d^3x\int d^3p\frac{1}{B}\epsilon^{ik}\delta^{jl}\Tr\Big[\frac{\partial Q_W(x,p)}{\partial p_{(i}}\star (D_W)_{j)}\star G_W(x,p)\star\\
&\frac{\partial Q_W(x,p)}{\partial p_{(k}}\star (D_W)_{l)}\star G_W(x,p)\star \frac{\partial Q_W(x,p)}{\partial \omega}\star G_W(x,p)\Big]\\
\nonumber =&\frac{1}{32\pi^2\beta_T A}\int d^3x\int d^3p\frac{1}{B}\epsilon^{ik}\delta^{jl}\Tr\Big[\frac{\partial Q_W(x,p)}{\partial p_i}\star (D_W)_j\star G_W(x,p)\star\\
&\frac{\partial Q_W(x,p)}{\partial p_k}\star (D_W)_l\star G_W(x,p)\star \frac{\partial Q_W(x,p)}{\partial \omega}\star G_W(x,p)\Big].
\label{generaltopologicalterm}
\end{align}
Round brackets on indices represent normalized symmetrization. Notice that this expression for the geometric Hall viscosity coefficient is only valid for the leading term in a filling factor expansion due to the obstruction produced by commutators of Weyl symbols with respect to the Moyal star product which is finite in the absence of Galilean boost symmetry. If the phase space dependence $(x,p)$ of the fermion operator and its inverse reduces to $p_{\mu}-A_{\mu}(x)$ (which comprises arbitrary electromagnetic fields in general), we will be able to replace exactly one Moyal star product in an integrand subject to boundaryless phase space integration by an ordinary product, since
\begin{align}
\nonumber \star =&{\rm exp} \Big(\frac{i}{2} \left( \overleftarrow{(\partial_x)}^{\mu}\overrightarrow{(\partial_p)}_{\mu}-\overleftarrow{(\partial_p)}_{\mu}\overrightarrow{(\partial_x)}^{\mu}\right) \Big)={\rm exp} \Big(\frac{i}{2} \left( \overleftarrow{(\partial_p)}^{\nu}((\partial_x)^{\mu}A_{\nu}(x))\overrightarrow{(\partial_p)}_{\mu}-\overleftarrow{(\partial_p)}_{\mu}((\partial_x)^{\mu}A_{\nu}(x))\overrightarrow{(\partial_p)}^{\nu}\right) \Big)\\
=&{\rm exp} \Big(\frac{i}{2} \left( \overleftarrow{(\partial_p)}^{\nu}F_{\mu\nu}(x)\overrightarrow{(\partial_p)}^{\mu}\right) \Big)={\rm exp} -\Big(\frac{i}{2} \left( \overrightarrow{(\partial_p)}^{\nu}F_{\mu\nu}(x)\overrightarrow{(\partial_p)}^{\mu}\right) \Big)=1.
\label{starproductreduction}
\end{align}
Importantly we employed integration by parts in the second line which is the reason why this identity allows for the substitution of one Moyal star product only. The expression in Eq. (\ref{generaltopologicalterm}) is immune to perturbative Coulomb interactions which is justified in detail in \cite{selch2026nonrenormalization}.\par
The expression for the Hall conductivity on the other hand may be simplified to ($p_3=\omega$)
\begin{align}
\sigma_H=\frac{1}{2\pi}\mathcal{N}_{\sigma},\,\,\,\,\mathcal{N}_{\sigma}=&{\frac{1}{24\pi^2\beta_T A}}\int d^3x \int d^3p\,\epsilon_{\mu\nu\rho}\Tr\Big[ \frac{\partial
Q_W(x,p)}{\partial p_{\mu}}\star G_W(x,p)\nonumber \\
&\star \frac{\delta Q_W(x,p)}{\delta p_{\nu}}\star G_W(x,p)\star \frac{\partial Q_W(x,p)}{\partial p_{\rho}}\star G_W(x,p)\Big]
\label{topologicalinvariant2}
\end{align}
using only the assumption $\frac{\partial}{\partial A_i}=-\frac{\partial}{\partial p_i}$ when acting on the fermion operator Weyl symbol $Q_W$. This expression is the first Chern number which is topological under general infinitesimal variations $Q_W\to Q_W+\delta Q_W$ as may be shown straightforwardly employing integration by parts for a boundaryless phase space. \par
We now come to relations of the derivatives of the fermion bilinear operator Weyl symbol for the case of Dirac fermions. We find
\begin{align}
&\frac{\partial Q_W}{\partial e_a^i}=\gamma^a(D_W)_i,\,\,\,\,-\frac{\partial Q_W}{\partial A_i}=\frac{\partial Q_W}{\partial p_i}=-i\gamma^ae_a^{i}\\
&\Rightarrow \frac{\partial Q_W}{\partial e_a^i}=i\frac{\partial Q_W}{\partial p_j}e_j^a\star (D_W)_i\Big\rvert_{e_a^i=\delta_a^i}=i (D_W)_i\star \frac{\partial Q_W}{\partial p_j}e_j^a\Big\rvert_{e_a^i=\delta_a^i}.
\label{vierbeintomomentum}
\end{align}
Together with the assumption $\frac{\partial}{\partial A_i}=-\frac{\partial}{\partial p_i}$ we may simplify the expressions for the geometric Hall viscosity and the Hall conductivity. In the former case we obtain
\begin{align}
\nonumber \eta_H^{geo}=\frac{1}{2\pi}\mathcal{N}_{\eta}B,\,\,\,\, \mathcal{N}_{\eta}=&\frac{1}{16\pi^2\beta_T A}\int d^3x\int d^3p\frac{\epsilon^{ik}\delta^{jl}}{B}\Tr\Big[\frac{\partial Q_W}{\partial p_i}\star (D_W)_j\star G_W(x,p)\\
&\star\frac{\partial Q_W}{\partial p_k}\star (D_W)_l\star G_W(x,p)\star\frac{\partial Q_W(x,p)}{\partial \omega}\star G_W(x,p)\Big].
\label{generaltopologicalterm2}
\end{align}
This expression coincides with Eq. (\ref{generaltopologicalterm}) structure-wise and deviates by an overall factor of two. This factor of two emerges due to the employed definitions of the geometric Hall viscosities for Pauli and Dirac fermions. In the literature the Hall viscosity of non-relativistic fermions is usually defined to be twice as large as our choice. The expression in Eq. (\ref{generaltopologicalterm2}) represents the geometric Hall viscosity coefficient faithfully only for systems respecting full Lorentz symmetry, as the relations in the second line of Eq. (\ref{vierbeintomomentum}) rely on this assumption. If $Q_W$ is an arbitrary function of the kinetic operator Weyl symbol $\gamma^ae_a^i(D_W)_i$ (for constant vielbein and in the absence of Lorentz boost symmetry), which is in analogy to considerations above, the Green function expression only represents the leading coefficient in a filling factor expansion correctly, again due to obstructions in form of commutators of Weyl symbols. Their appearance and relevance may be shown in analogy to the non-relativistic case. A counterexample where the relations in Eq. (\ref{vierbeintomomentum}) are not valid and consequently the Green function expression in Eq. (\ref{generaltopologicalterm2}) does not faithfully represent the geometric Hall viscosity coefficient is biased Bernal bilayer graphene in the low energy chiral fermion approximation as alluded to in the main text. The coefficient $\mathcal{N}_{\eta}$ is only topological in a restricted sense in complete analogy to the non-relativistic case.\par
The Hall conductivity for Dirac fermions is identical in structure as compared to the previous case of Pauli fermions
\begin{align}
\sigma_H=\frac{1}{2\pi}\mathcal{N}_{\sigma},\,\,\,\,\mathcal{N}_{\sigma}=&{\frac{1}{24\pi^2\beta_T A}}\int d^3x \int d^3p\,\epsilon_{\mu\nu\rho}\Tr\Big[ \frac{\partial
Q_W(x,p)}{\partial p_{\mu}}\star G_W(x,p)\nonumber \\
&\star \frac{\delta Q_W(x,p)}{\delta p_{\nu}}\star G_W(x,p)\star \frac{\partial Q_W(x,p)}{\partial p_{\rho}}\star G_W(x,p)\Big].
\label{topologicalinvariant4}
\end{align}
The topological invariance of the Hall conductivity of Dirac fermions may be proven in a way identical to Pauli fermions. 
Note that the Moyal star products may not be removed in the presence of a homogeneous magnetic field, since not all canonical momenta are good quantum numbers. Instead we find
\begin{align}
\star =\exp \Big(i\frac{B}{2}\epsilon_{ij}\overset{\leftarrow}{\partial}_{(D_W)_i}\overset{\rightarrow}{\partial}_{(D_W)_j}\Big).
\end{align}
We will calculate the topological coefficients for Dirac and Pauli fermions and discuss the result in the next section. The corresponding coefficients for Pauli fermions were discussed previously in detail in \cite{qhemaik2,selch2026nonrenormalization}. We will follow the same steps for Dirac fermions. 

\section{Appendix J: Calculation of the emergent Hall viscosity}

In this appendix we detail the derivation of the emergent Hall viscosity. We will show that the calculation splits into three independent contributions each of which will then be discussed consecutively.

We start from the stress tensor as derived from the strain tensor and apply the chain rule to extract the dependence of emergent strain gauge field and metric on strain. We obtain
\begin{align}
\nonumber \bar{T}_{ij}^{str}(x)=&\frac{1}{\beta_T A}\int d^3x\Big(-\frac{1}{Z}\frac{\delta Z}{\delta u_{ij}(x)}\Big)=\frac{1}{\beta_T A}\int d^3xd^3y\Big(-\frac{1}{Z}\frac{\delta Z}{\delta A^s_m(y)}\frac{\delta A^s_m(y)}{\delta u_{ij}(x)}-\frac{1}{Z}\frac{\delta Z}{\delta e_a^m(y)}\frac{\delta e_a^m(y)}{\delta u_{ij}(x)}\Big)\\
\nonumber =&\frac{1}{\beta_T A}\int d^3x\frac{d^3p}{(2\pi )^3}\Big(\Tr\Big[\frac{\delta Q_W}{\delta A^s_m(x)}\frac{\partial A^s_m}{\partial u_{ij}}\star G_W\Big]+\Tr\Big[\frac{\delta Q_W}{\delta e_a^m(x)}\frac{\partial e_a^m}{\partial u_{ij}}\star G_W\Big]\Big)\\
\nonumber =&-\frac{1}{2\beta_T A}\int d^3x\frac{d^3p}{(2\pi )^3}\Big(\Tr\Big[\frac{\partial Q_W}{\partial A^s_m}\star G_W\star\frac{\partial Q_W}{\partial [A^s_n}\star G_W\star \frac{\partial Q_W}{\partial \omega]}\star G_W\Big]\frac{\partial A^s_m}{\partial u_{ij}}\frac{\partial A^s_n}{\partial u_{kl}}\\
\nonumber &+\Tr\Big[\frac{\partial Q_W}{\partial A^s_m}\star G_W\star\frac{\partial Q_W}{\partial [e_a^n}\star G_W\star \frac{\partial Q_W}{\partial \omega]}\star G_W\Big]\frac{\partial A^s_m}{\partial u_{ij}}\frac{\partial e_a^n}{\partial u_{kl}}\\
\nonumber &+\Tr\Big[\frac{\partial Q_W}{\partial e_a^m}\star G_W\star\frac{\partial Q_W}{\partial [A^s_n}\star G_W\star \frac{\partial Q_W}{\partial \omega]}\star G_W\Big]\frac{\partial e_a^m}{\partial u_{ij}}\frac{\partial A^s_n}{\partial u_{kl}}\\
&+\Tr\Big[\frac{\partial Q_W}{\partial e_a^m}\star G_W\star\frac{\partial Q_W}{\partial [e_b^n}\star G_W\star \frac{\partial Q_W}{\partial \omega]}\star G_W\Big]\frac{\partial e_a^m}{\partial u_{ij}}\frac{\partial e_b^n}{\partial u_{kl}}\Big)\partial_0u_{kl}.
\end{align}
Square brackets on variables represent antisymmetrization. The partial differentiations of the fermion bilinear operator may be simplified as follows. By translational invariance modulo a vector potential representing the constant magnetic background field we find a reduced phase space dependence. Both the bilinear fermion operator and its inverse are functions of the covariant derivative $(D_W)_i=-ip_{\mu}+iA_{\mu}$. This implies $\frac{\partial}{\partial A_i}=-\frac{\partial}{\partial p_i}$. Analogously the differentiation with respect to the vielbein may be simplified. We find
\begin{align}
&\frac{\partial Q_W}{\partial e_a^i}=\gamma^a(D_W)_i,\,\,\,\,-\frac{\partial Q_W}{\partial A_i}=\frac{\partial Q_W}{\partial p_i}=-i\gamma^ae_a^{i}\\
&\Rightarrow \frac{\partial Q_W}{\partial e_a^i}=i\frac{\partial Q_W}{\partial p_j}e_j^a\star (D_W)_i\Big\rvert_{e_a^i=\delta_a^i}=i (D_W)_i\star \frac{\partial Q_W}{\partial p_j}e_j^a\Big\rvert_{e_a^i=\delta_a^i}.
\end{align}
The emergent Hall viscosity $\eta_H^{em}$ is defined by
\begin{align}
\eta_H^{em}=\frac{1}{4}\epsilon^{ik}\delta^{jl}\frac{\partial \bar{T}_{ij}^{str}}{\partial \partial_0u_{kl}}.
\end{align}
Now we simplify the contracted differentials of the emergent gauge field and metric relevant to the emergent Hall viscosity with respect to the strain field. This leads to the index structures
\begin{align}
&\frac{1}{4}\epsilon^{ik}\delta^{jl}\frac{\partial A^s_m}{\partial u_{ij}}\frac{\partial A^s_n}{\partial u_{kl}}=\frac{\beta^2}{16a^2}\epsilon^{ik}\delta^{jl}K_{mip}K_{nkr}\epsilon_{pj}\epsilon_{rl}=-\frac{\beta^2}{8a^2}\epsilon^{mn}\\
&\frac{1}{4}\epsilon^{ik}\delta^{jl}\frac{\partial A^s_m}{\partial u_{ij}}\frac{\partial e_a^n}{\partial u_{kl}}=\frac{\beta (1-\beta )}{16a}\epsilon^{ik}\delta^{jl}K_{mip}\epsilon_{pj}\Big(\delta_a^k\delta^{ln} +\delta_a^l\delta^{kn} +(\frac{\beta}{1-\beta}\delta^{kl}\delta^n_a)\Big)\nonumber\\&=\frac{\beta (1-\beta )}{16a}(\epsilon^{ia}K_{mip}\epsilon_{pn}+\epsilon^{in}K_{mip}\epsilon_{pa})\\
&\frac{1}{4}\epsilon^{ik}\delta^{jl}\frac{\partial e_a^m}{\partial u_{ij}}\frac{\partial A^s_n}{\partial u_{kl}}=\frac{\beta (1-\beta )}{16a}\epsilon^{ik}\delta^{jl}\Big(\delta_a^i\delta^{jm} +\delta_a^j\delta^{im}+(\frac{\beta}{1-\beta}\delta^{ij}\delta^m_a)\Big)K_{nkp}\epsilon_{pl}\nonumber\\&
=\frac{\beta (1-\beta )}{16a}(\epsilon^{ak}K_{nkp}\epsilon_{pm}+\epsilon^{mk}K_{nkp}\epsilon_{pa})\\
&\frac{1}{4}\epsilon^{ik}\delta^{jl}\frac{\partial e_a^m}{\partial u_{ij}}\frac{\partial e_b^n}{\partial u_{kl}}=\frac{(1-\beta )^2}{16}\epsilon^{ik}\delta^{jl}\Big(\delta_a^i\delta^{jm} +\delta_a^j\delta^{im}+(\frac{\beta}{1-\beta}\delta^{ij}\delta^m_a)\Big)\Big(\delta_b^k\delta^{ln} +\delta_b^l\delta^{kn}+(\frac{\beta}{1-\beta}\delta^{kl}\delta^n_b)\Big)\nonumber\\&=\frac{(1-\beta )^2}{16}(\delta^{mn}\epsilon_{ab}+\delta^{bm}\epsilon_{an}+\delta^{an}\epsilon_{mb}+\delta^{ab}\epsilon_{mn}).
\end{align}
The emergent Hall viscosity may finally be represented by
\begin{align}
\nonumber \eta_H^{em}=\frac{1}{2\pi}\Big[&\Big(\frac{\beta}{2a}\Big)^2\mathcal{N}_{\sigma}+\frac{\beta (1-\beta )}{2a}(\mathcal{M}_1+\mathcal{M}_2)\sqrt{B}+(1-\beta )^2\mathcal{N}_{\eta}B\Big]
\end{align}
with ($\omega =p_3$)
\begin{align}
\mathcal{N}_{\sigma}=&\frac{1}{24\pi^2\beta_T A}\int d^3xd^3p\epsilon_{\mu\nu\rho}\Tr\Big[\frac{\partial Q_W}{\partial p_{\mu}}\star G_W\star\frac{\partial Q_W}{\partial p_{\nu}}\star G_W\star \frac{\partial Q_W}{\partial p_{\rho}}\star G_W\Big]\\
\mathcal{M}_1=&\frac{{i}}{64\pi^2\beta_T A}\int d^3xd^3p\frac{(\epsilon^{ia}K_{mip}\epsilon_{pn}+\epsilon^{in}K_{mip}\epsilon_{pa})}{\sqrt{B}}\Tr\Bigl[\frac{\partial Q_W}{\partial p_m}\star G_W\star\frac{\partial Q_W}{\partial p_a}\star (D_W)_n\star G_W\star \frac{\partial Q_W}{\partial p_3}\star G_W\nonumber\\ & - (p_m \leftrightarrow p_3)\Bigr]\\
\mathcal{M}_2=&\frac{{i}}{64\pi^2\beta_T A}\int d^3xd^3p\frac{(\epsilon^{ak}K_{nkp}\epsilon_{pm}+\epsilon^{mk}K_{nkp}\epsilon_{pa})}{\sqrt{B}}\Tr\Bigl[\frac{\partial Q_W}{\partial p_a}\star (D_W)_m\star G_W\star\frac{\partial Q_W}{\partial p_n}\star G_W\star \frac{\partial Q_W}{\partial p_3}\star G_W\nonumber\\& - (p_n \leftrightarrow p_3)\Bigr]\\
\mathcal{N}_{\eta}=&\frac{1}{128\pi^2\beta_T A}\int d^3xd^3p\frac{(\delta^{mn}\epsilon_{ab}+\delta^{bm}\epsilon_{an}+\delta^{an}\epsilon_{mb}+\delta^{ab}\epsilon_{mn})}{B}\cdot \nonumber\\& \Tr\Big[\frac{\partial Q_W}{\partial p_a}\star (D_W)_m\star G_W\star\frac{\partial Q_W}{\partial p_b}\star (D_W)_n\star G_W\star \frac{\partial Q_W}{\partial p_3}\star G_W - ((a,m)\leftrightarrow (b,n))\Big]\nonumber\\
=&\frac{1}{16\pi^2\beta_T A}\int d^3xd^3p\frac{\delta^{mn}\epsilon_{ab}}{B}\Tr\Big[\frac{\partial Q_W}{\partial p_a}\star (D_W)_m\star G_W\star\frac{\partial Q_W}{\partial p_b}\star (D_W)_n\star G_W\star \frac{\partial Q_W}{\partial p_3}\star G_W\Big].
\end{align}
The coefficient $\mathcal{N}_{\sigma}$ represents the Hall conductivity Chern number while $\mathcal{N}_{\eta}$ represents the Hall viscosity topological invariant which is one fourth of the product of Wen-Zee shift times the Hall conductivity Chern number. The ``mixed'' coefficients $\mathcal{M}_i$ ($i=1,2$) are naturally related to the physics of piezoelectricity in general which is not well defined in the integer quantum Hall regime due to its incompressibility, though. 

\section{Appendix K: Expression for $\mathcal{N}_{\eta}$ and $\mathcal{N}_{\sigma}$ through the effective Hamiltonian}

We proceed now to calculate the coefficients $\mathcal{N}_{\sigma}$, $\mathcal{M}_i$ ($i=1,2$) and $\mathcal{N}_{\eta}$ explicitly in the integer quantum Hall regime. The calculation closely follows those presented in \cite{qhemaik2,selch2026nonrenormalization}. We start with the Hall viscosity coefficient $\mathcal{N}_{\eta}$ for Dirac fermions and notice that the other coefficients may be calculated analogously just removing covariant derivative terms in the expression of $\mathcal{N}_{\eta}$ and reorganizing index contractions. The necessary adjustments will be pointed out along the way. We evaluate $\mathcal{N}_{\eta}$ with the normalization coinciding with the standard one employed in the literature (which implies $g_{kl}\to g_{kl}/2$ in the definitions in Eq. (\ref{geodef}) for Pauli fermions).\par
For definiteness we choose the Landau gauge in which the vector potential $\bold{A}$ is given by
\begin{align}
A_x = 0, \quad A_y = B x\,. 
\end{align}
We will start from the expression in Eq. (\ref{generaltopologicalterm})
\begin{align}  \label{N-2}
\nonumber \mathcal{N}_{\eta}=&{\frac{T}{16\pi^2 A}}\int d^3p d^3x \frac{\delta^{lm}}{B}\epsilon_{ij}\Tr\Big[ G_W\star \partial_{p_3} Q_W\star G_W\star\\
&\partial_{p_i} Q_W\star (D_W)_l\star G_W \star \partial_{p_j} Q_W\star (D_W)_m \Big]. 
\end{align}
In Wigner representation, the quantity $\partial_{p_i} Q_W$ is calculated as follows
\begin{align}
\nonumber \frac{\partial}{\partial p^i}Q_W(x,p)=&\frac{\partial}{\partial p^i}\int \frac{d^3P}{(2\pi )^3}~e^{iPx} \tilde{Q}\Big(p+\frac{P}{2},p-\frac{P}{2}\Big)\\
=&\int \frac{d^3P}{(2\pi )^3}~e^{iPx}\Big(\frac{\partial}{\partial K_1^i }+\frac{\partial}{%
\partial {K_2^i} }) \tilde{Q}(K_1,K_2)\Big|_{K_1=p+\frac{P}{2}}^{K_2=p-\frac{P%
}{2}}
\end{align}
where we used the chain rule of differentiation to get 
\begin{align}
\frac{\partial}{\partial {p^i} } =\frac{\partial K_1^j}{\partial {p^i} }%
\frac{\partial}{\partial {K_1^j} }+\frac{\partial K_2^j}{\partial {p^i} }%
\frac{\partial}{\partial {K_2^j} } =\frac{\partial}{\partial {K_1^i} }+\frac{%
\partial}{\partial {K_2^i} }.
\end{align}
If we denote 
\begin{align}
\tilde{Q}_i\Big(p+\frac{P}{2},p-\frac{P}{2}\Big)= \Big( \frac{\partial}{\partial {K_1^i}} +\frac{\partial}{\partial{K_2^i}}\Big) 
\tilde{Q}(K_1,K_2)\Big|_{K_1=p+P/2}^{K_2=p- P/2}
\end{align}
then $\partial_{p_i}Q_W$ can be written in short-hand form as $\partial_{p_i}Q_W=(Q_{eff,i})_W$. Using the associativity of the Moyal star product Eq. (\ref{N-2}) is written as 
\begin{align}
\nonumber &G_W \star \partial_{p_3} Q_W \star G_W\star\partial_{p_i} Q_W\star (D_W)_l \star G_W \star \partial_{p_j} Q_W\star (D_W)_m\\
&=\int \frac{d^3P}{(2\pi )^3}~e^{iPx}(\tilde{G}\tilde{Q}_3\tilde{G}\tilde{Q}_j\tilde{D}_l\tilde{G}\tilde{Q}_{eff,k}\tilde{D}_m)
\Big(p+\frac{P}{2},p-\frac{P}{2}\Big)
\end{align}
where 
\begin{eqnarray}
&& (\tilde{G}\tilde{Q}_3\tilde{G}\tilde{Q}_i\tilde{D}_l\tilde{G}\tilde{Q}_j\tilde{D}_m)\Big(p+\frac{P}{2},p-\frac{P}{2}\Big) = \nonumber \\
&&  \int d^3p^{(2)}d^3p^{(3)}d^3p^{(4)}d^3p^{(5)}d^3p^{(6)}d^3p^{(7)}\delta^3\Big(p^{(7)}-p+\frac{P}{2}\Big)\cdot  \nonumber \\
&& \Big[ \tilde{G}(p + \frac{P}{2},p^{(2)})\Big( \lbrack\partial_{p^{(2)}_3} + \partial_{p^{(3)}_3}] \tilde{Q}(p^{(2)},p^{(3)})\Big)\cdot \nonumber \\
&& \tilde{G}(p^{(3)},p^{(4)})\Big( \lbrack \partial_{p^{(4)}_i} +\partial_{p^{(5)}_i}] \tilde{Q}(p^{(4)},p^{(5)})\tilde{D}_l\Big)\cdot  \nonumber \\
&& \tilde{G}(p^{(5)},p^{(6)})\Big( \lbrack \partial_{p^{(6)}_j} +\partial_{p^{(7)}_j}] \tilde{Q}(p^{(6)},p^{(7)})\tilde{D}_m\Big) \Big].  
\label{GQ0GQiGQj}
\end{eqnarray}
Substituting the above expressions back into Eq. (\ref{N-2}) and integrating over $x$ and $P$ leads to ($T$ vanishes due to imaginary time independence of the propagator (and its inverse) and integration over $d\tau$)
\begin{align}
\mathcal{N}_{\eta}=&{\frac{1}{16\pi^2A}} \int d^3p^{(1)}d^3p^{(2)}d^3p^{(3)}d^3p^{(4)}d^3p^{(5)}
d^3p^{(6)}\cdot  \nonumber \\
&\frac{\delta^{lm}\epsilon_{ij}}{B}\Tr\Big[\tilde{G}(p^{(1)} ,p^{(2)})\Big( \lbrack\partial_{p^{(2)}_3} + \partial_{p^{(3)}_3}] \tilde{Q}(p^{(2)},p^{(3)})\Big)\cdot \nonumber \\
&\tilde{G}(p^{(3)},p^{(4)})\Big( \lbrack \partial_{p^{(4)}_i} +\partial_{p^{(5)}_i}] \tilde{Q}(p^{(4)},p^{(5)})\tilde{D}_l\Big)\cdot  \nonumber \\
&\tilde{G}(p^{(5)},p^{(6)})\Big( \lbrack \partial_{p^{(6)}_j} +\partial_{p^{(1)}_j}] \tilde{Q}(p^{(6)},p^{(1)})\tilde{D}_m\Big)\Big].
\label{GQ0GQiGQj-2}
\end{align}
The momentum space function $\tilde{Q}(p^{(i)},p^{(j)})$ has the representation 
\begin{align}
\tilde{Q}(p^{(i)},p^{(j)}) = \langle p^{(i)}| \hat{Q} | p^{(j)}\rangle =\Big(  i\omega^{(i)}\delta^{2} (p^{(i)} - p^{(j)}) - \langle p^{(i)} |
\hat{H} | p^{(j)} \rangle \Big) \delta (\omega^{(i)}-\omega^{(j)})  \label{Q1}
\end{align}
where $p = (p_3,p_1,p_2) = (\omega, \mathbf{p})$. The corresponding Green function can be calculated as 
\begin{align}
\tilde{G}(p^{(i)},p^{(j)})= \sum_{n} \frac{1}{i\omega^{(i)} - \mathcal{E}_n}
\langle {p}^{(i)}| n \rangle \langle n | p^{(j)}\rangle
\delta(\omega^{(i)}-\omega^{(j)})\,. 
\end{align}
The numbers $\mathcal{E}_n$ represent the energy eigenvalues of $\hat{H}$ shifted by the chemical potential $\mu$. From Eq. (\ref{Q1}) it may be concluded that 
\begin{align}
\partial_{p^{(i)}_3} \tilde{Q} (p^{(i)},p^{(i+1)})=& i\delta^{2} (p^{(i)}- p^{(i+1)})\delta (\omega^{(i)}-\omega^{(i+1)}), \label{p0' Q}\\
\nonumber (\partial_{p^{(i)}_j} + \partial_{p^{(i+1)}_j}) \tilde{Q} (p^{(i)},p^{(i+1)})=& -(\partial_{p^{(i)}_j} + \partial_{p^{(i+1)}_j}) \langle p^{(i)} |
\hat{H} | p^{(j)} \rangle \delta(\omega^{(i)}-\omega^{(i+1)})  \\
\nonumber =& -i \langle p^{(i)} | \hat{H} \hat x_j - \hat x_j \hat{H} | p^{(i+1)}\rangle \delta(\omega^{(i)}-\omega^{(i+1)}) \\
=& -i \langle p^{(i)} | [\hat{H}, x_j]| p^{(i+1)}\rangle \delta(\omega^{(i)}-\omega^{(i+1)}).
\label{pro-Q}
\end{align}
We can easily integrate out all intermediate $\omega^{(i)}$'s by means of the delta functions to remain with a single $\omega$-integration and then proceed to integrate
out $(p^{(2)},p^{(3)})$ in Eq. (\ref{GQ0GQiGQj-2}) after substituting the expression of Eq. (\ref
{p0' Q}) and using the completeness relation $\hat{1}=\int d^3p_i|p_i \rangle \langle p_i |$. After such operations Eq. (\ref{GQ0GQiGQj-2}) reduces to 
\begin{align}
\mathcal{N}_{\eta}=&\frac{i}{16\pi^2A}\,\sum_{n,k} \int d\omega d^2p^{(1)}d^2p^{(2)}d^2p^{(3)}d^2p^{(4)}\frac{\delta^{lm}}{B} \epsilon_{ij}\cdot  \nonumber\\
\nonumber &\mathrm{Tr}\,\Big[ \frac{1}{(i\omega^{} - \mathcal{E}_n)^2} \langle {\ p}^{(1)}| n \rangle \langle n | {\ p}^{(2)}\rangle \langle {p}^{(2)}| [\hat{H}, \hat x_i]\tilde{D}_l | {\ p}^{(3)}\rangle\cdot\\
&\frac{1}{(i\omega^{} - \mathcal{E}_k)}\langle {p}^{(3)}| k \rangle \langle k | {\ p}^{(4)}\rangle \langle {\ p}
^{(4)}| [\hat{H}, \hat x_j] \tilde{D}_m| {p}^{(1)}\rangle \Big]
\end{align}
where we redefine the momentum variables left as $p^{(1)}, p^{(2)},p^{(3)}, p^{(4)}$. Further integrating out the momenta $p^{(i)}$ we have 
\begin{align}  
\label{N-3}
\nonumber \mathcal{N}_{\eta} =& \frac{i}{16\pi^2A}\,\sum_{n,k} \int \,d \omega d^2p\frac{\delta^{lm}}{B}
\epsilon_{ij}\,\Big[ \frac{1}{(i\omega^{} - \mathcal{E}_n)^2}\langle n| [\hat{H}, {\hat x}_i]\tilde{D}_l | k \rangle \frac{1}{(i\omega^{} - \mathcal{E}_k)}
\langle k | [\hat{H}, {\hat x}_j]\tilde{D}_m | n \rangle \Big] \\
=& -\frac{i}{8\pi A}\,\sum_{n,k}\frac{\delta^{lm}}{B} \epsilon_{ij}\, \langle n | [\hat{
H}, \hat x_i]\tilde{D}_l | k \rangle \langle k | [\hat{H}, \hat x_j] \tilde{D}_n| n
\rangle\frac{ (\Theta (-\mathcal{E}_n) \Theta(\mathcal{E}_k) - \Theta (-
\mathcal{E}_k) \Theta(\mathcal{E}_n))}{(\mathcal{E}_k - \mathcal{E}_n)^2}. 
\end{align}
In order to obtain the second equality in Eq. (\ref{N-3}) we have employed the residue theorem and compactified the resulting expression using Heaviside step functions $\Theta$.\par
Now, we will define new variables $(\xi_i, X_i)$ as follows 
\begin{align}
\nonumber &\hat{x}_1 = -\frac{\hat{p}_y - B x}{B} + \hat{X}_1 = \hat{\xi}_1+ \hat{X}_1,\\
&\hat{x}_2 = \frac{\hat{p}_x}{B} + \hat{X}_2= \hat{\xi}_2 + \hat{X}_2,\,\,\,\,-iD_l=B\epsilon_{lm}\hat{\xi}_m. 
\end{align}
The commutation relations implied by the above set of variables are 
\begin{align}
[\hat{\xi}_{i},\hat{\xi}_{j}] = \frac{i}{B}\epsilon_{ij}, \,\,\,\, [\hat{X}_{i},\hat{X}_{j}] = - \frac{i}{B}\epsilon_{ij},\,\,\,\,[\hat{H}, \hat{X}_{i}] = 0 \,\,\,\,\Big(\hat{H}=\frac{B^2}{2m}(\hat{\xi}_1^2+\hat{\xi}_2^2)\Big)
\end{align}
where $i$ labels the first and second set of variables, respectively. In terms of the $\hat{\xi}_i$'s, Eq. (\ref{N-3}) can be written as 
\begin{align}
A\mathcal{N}_{\eta} =\frac{iB^2}{8\pi} \sum_{n,k} \epsilon_{ij}\frac{\delta^{lm}}{B} \,\Big[ \langle n| [\hat{H}
, {\hat \xi}_i] \hat{\xi}_l| k \rangle\langle k | [\hat{H}, {\hat \xi}_j]\hat{\xi}_m | n
\rangle \Big] \frac{( \Theta (-\mathcal{E}_n)\Theta(\mathcal{E}_k) - \Theta (-
\mathcal{E}_k)\Theta(\mathcal{E}_n))}{(\mathcal{E}_k - \mathcal{E}_n)^2}.
\end{align}
The two terms involving Heaviside step functions entering with opposite signs give identical contributions to $A\mathcal{N}_{\eta_H}$ which may be checked easily using Eqs. (\ref{matrixelements}) and (\ref{matrixelements2}) below. We may then insert two further resolutions of the identity in terms of energy eigenstates to write
\begin{align}
\nonumber A\mathcal{N}_{\eta} =&\frac{iB^2}{4\pi} \sum_{n,k,q,r} \epsilon_{ij}\frac{\delta^{lm}}{B} \,\Big[ \langle n| [\hat{H}
, {\hat \xi}_i]|q\rangle \langle q| \hat{\xi}_l| k \rangle\langle k | [\hat{H}, {\hat \xi}_j]|r\rangle \langle r|\hat{\xi}_m | n
\rangle \Big] \frac{\Theta (-\mathcal{E}_n)\Theta(\mathcal{E}_k)}{(\mathcal{E}_k - \mathcal{E}_n)^2}\\
=&\frac{iB^2}{4\pi} \sum_{n,k,q,r} \epsilon_{ij}\frac{\delta^{lm}}{B} \,\Big[ \langle n|{\hat \xi}_i|q\rangle \langle q| \hat{\xi}_l| k \rangle \langle k | {\hat \xi}_j|r\rangle \langle r|\hat{\xi}_m | n
\rangle \Big]\frac{(\mathcal{E}_n-\mathcal{E}_q)\cdot (\mathcal{E}_k-\mathcal{E}_r)}{(\mathcal{E}_k - \mathcal{E}_n)^2}\Theta (-\mathcal{E}_n)\Theta(\mathcal{E}_k).
\label{hallviscosityintermediate}
\end{align}
The indices $n$, $k$, $q$, $r$ have a discrete index labeling Landau levels and a continuous index labeling intra Landau level states. We may now employ the relations for matrix elements of the effective Pauli Landau level Hamiltonian assuming for simplicity that $A=L^2$ is square shaped. In the language of appendix B1 the Dirac ket $|n\rangle$ up to now refers to the eigenfunction $\Psi_n$. In the following matrix element relations we employ Pauli kets which are written identically as compared to Dirac kets but are meant to refer to the eigenfunctions $\phi_n$
\begin{align}
&\langle n|\hat{\xi}_1|k\rangle =\frac{l_B}{L}\delta (n_{con}-k_{con})\Big(\sqrt{\frac{n_{dis}}{2}}\delta_{n_{dis},k_{dis}+1}+\sqrt{\frac{n_{dis}+1}{2}}\delta_{n_{dis},k_{dis}-1}\Big),\label{matrixelements}\\
&\langle n|\hat{\xi}_2|k\rangle =i\frac{l_B}{L}\delta (n_{con}-k_{con})\Big(\sqrt{\frac{n_{dis}}{2}}\delta_{n_{dis},k_{dis}+1}-\sqrt{\frac{n_{dis}+1}{2}}\delta_{n_{dis},k_{dis}-1}\Big).
\label{matrixelements2}
\end{align}
The quantity $l_B$ is known as the magnetic length and parametrizes the characteristic extension of electron wavefunctions (see, e. g., \cite{Tong:2016kpv}). We assume that in the field theory $p-1$ effective Landau levels above the unpaired Landau level are fully filled. Furthermore our momentum space scaling and the Landau level degeneracy imply the intra Landau level momentum range
\begin{align}
\Delta p_y=2\pi \frac{L}{l_{B}^2},\,\,\,\, l_{B}^2=B^{-1}.
\label{momentumrange}
\end{align}
Notice that the matrix elements in Eqs. (\ref{matrixelements}) and (\ref{matrixelements2}) are nonzero only for eigenstates corresponding to adjacent eigenvalues in the Landau level spectrum. Together with the contraints implied by the Heaviside functions we find that $k_{dis}=p,p+1$ and $n_{dis}=q_{dis}-1=r_{dis}-1=k_{dis}-2$. 
Fully employing both Eqs. (\ref{matrixelements}), (\ref{matrixelements2}) and (\ref{momentumrange}) within Eq. (\ref{hallviscosityintermediate}) then leads to $\mathcal{N}_{\eta}=\frac{1}{8}(p^2+(p-1)^2)$ per valley and per spin degree of freedom for the Hall viscosity coefficient of massless Dirac fermions. The explicit calculation involves mixing of states which are particle-hole conjugate as well as the zero Landau level and has been performed using symbolic python. Generalized massless Dirac fermions with arbitrary function $C$ and vanishing function $M$ yield the same result and imply the topological robustness of the massless case. A similar calculation for massive Dirac fermions and even generalized massive Dirac fermions may be performed similarly. In the case of (generalized) massive Dirac fermions it is found that the Hall viscosity coefficient acquires an explicit mass dependence and ceases from being topological.\par
An analogous calculation for the Hall conductivity coefficient $\mathcal{N}_{\sigma}$ for Dirac fermions up to an expression corresponding to Eq. (\ref{hallviscosityintermediate}) results in
\begin{align}
A\mathcal{N}_{\sigma}&=-\frac{i}{2\pi}
\sum_{n,k}\epsilon_{ij}\langle n| [\hat{H},\hat{\xi}_i]| k\rangle\langle
k|[\hat{H},\hat{\xi}_j]|n\rangle \frac{\Theta (-\mathcal{E}_n)\Theta (\mathcal{E}_k)}{(\mathcal{E}_n-\mathcal{E}_k)^2}  \nonumber \\
&=-\frac{i}{2\pi} \sum_{n,k}\epsilon_{ij}\langle n|\hat{\xi}_i| k\rangle\langle k|\hat{\xi}_j|n\rangle \Theta (-\mathcal{E}_n)\Theta (\mathcal{E}_k)  \nonumber\\
&=-\frac{i}{2\pi} \sum_{n,k}\epsilon_{ij}\langle n|\hat{\xi}_i| k\rangle\langle k|\hat{\xi}_j|n\rangle \Theta (-\mathcal{E}_n)(\Theta (\mathcal{E}_k)+\Theta (-\mathcal{E}_k))  \nonumber \\
&=-\frac{i}{2\pi} \sum_{n,k}\epsilon_{ij}\langle n|\hat{\xi}_i| k\rangle\langle k|\hat{\xi}_j|n\rangle \Theta (-\mathcal{E}_n)\nonumber\\
&=-\frac{i}{2\pi} \sum_{n}\langle n|[\hat{\xi}_1,\hat{\xi}_2]|n\rangle \Theta (-\mathcal{E}_n)=-\frac{i}{2\pi} \sum_{n}\langle n|\frac{i}{B}|n\rangle \Theta (-\mathcal{E}_n)\nonumber \\
&=\frac{1}{2\pi}\int dp_{y}\frac{1}{B}L\sum_{m}\Theta (-\mathcal{E}_m)=A\sum_{m}\Theta (-\mathcal{E}_m).
\end{align}
Since the energy spectrum of graphene is not bounded from below, this result requires regularization. A reference state is the unpaired Landau level which provides a finite Hall conductivity once the chemical potential is just below or above its energy eigenvalue. Then for $p-1$ filled effective Landau levels above the unpaired Landau level we obtain $\mathcal{N}_{\sigma}=p+(p-1)$ per spin degree of freedom. Notice that in this case we do not need to know explicitly the values of matrix elements of the form $\langle n|\hat{\xi}_i| k\rangle$.\par
The Pauli fermions emerging from the non-relativistic limit of massive Dirac fermions in gapped graphene have a valley dependent dispersion and therefore valley dependent quantized Landau level energy eigenvalues in a magnetic field according to Eqs. (\ref{nonrelativisticdispersion}) and (\ref{nonrelativisticeigenvalues}), respectively. The Hall conductivity and Hall viscosity as determined from the non-relativistic limit may be directly compared to the relativistic case. We find that if the chemical potential is in the gap and closer to the first unoccupied band than to the first occupied band and the first $p-1$ Landau levels above the unpaired Dirac Landau level are occupied then $\mathcal{N}_{\sigma}=p+(p-1)$ and $\mathcal{N}_{\eta}=\frac{1}{4}(p^2+(p-1)^2)$ per spin degree of freedom. This coincides with the relativistic case.\par 
The piezoelectric or mixed coefficients $\mathcal{M}_i$ ($i=1,2$) vanish identically both for Dirac and Pauli fermions. By comparison with Eqs. (\ref{hallviscosityintermediate}) it follows that these coefficients may be determined from sums over products of three matrix elements of the form $\langle n|\hat{\xi}_i| k\rangle$ with adjacent discrete indices $n$ and $k$. It may be seen straightforwardly that any expression comprising and odd number of such products must necessarily vanish as this criterion can not be fulfilled within a trace structure.\par
While $\mathcal{N}_{\sigma}$ counts the number of fully occupied Landau levels modulo the parity anomaly for Dirac fermions, $\mathcal{N}_{\eta}$ is found to be proportional to a quantized orbital spin of the involved fermions within the Hall fluid in the case of translational and rotational invariance (see, e. g., \cite{hoyos2014hall}).


\begin{thebibliography}{10}


\bibitem{castroneto2009the}
A. H. Castro Neto, F. Guinea, N. M. R. Peres, K. S. Novoselov and A. K. Geim.
\newblock {\em Rev. Mod. Phys.} {\bfseries 81}, 109 (2009).

\bibitem{goerbig2011electronic}
M. O. Goerbig.
\newblock {\em Rev. Mod. Phys.} {\bfseries 83}, 1193 (2011).

\bibitem{semenoff1984condensed}
G. W. Semenoff.
\newblock {\em Phys. Rev. Lett.} {\bfseries 53}, 2449 (1984).

\bibitem{abanov2014electromagnetic}
A. G. Abanov and A. Gromov.
\newblock {\ttfamily arXiv:1401.3703}.

\bibitem{gromov2014density}
A. Gromov and A. G. Abanov.
\newblock {\em Phys. Rev. Lett.} {\bfseries 113}, 266802 (2014).

\bibitem{hoyos2012hall}
C. Hoyos and D. T. Son.
\newblock {\em Phys. Rev. Lett.} {\bfseries 108}, 066805 (2012).

\bibitem{cho2014geometry}
G. Y. Cho, Y. You and E. Fradkin.
\newblock {\em Phys. Rev. B} {\bfseries 90}, 115139 (2014).






\bibitem{avron1995viscosity}
J. E. Avron, R. Seiler, and P. G. Zograf.
\newblock {\em Phys. Rev. Lett.} {\bfseries 75}, 697 (1995).

\bibitem{avron1998odd}
J. E. Avron.
\newblock {\em J. Stat. Phys.} {\bfseries 92}, 543 (1998).

\bibitem{read2009non}
N. Read.
\newblock {\em Phys. Rev. B} {\bfseries 79}, 045308 (2009).

\bibitem{hoyos2014hall}
C. Hoyos.
\newblock {\em Int. Jour. Mod. Phys. B} {\bfseries 28}, 1430007 (2014).

\bibitem{delacretaz2017transport}
L. V. Delacr\'etaz and A. Gromov.
\newblock {\em Phys. Rev. Lett.} {\bfseries 119}, 226602 (2017).

\bibitem{scaffidi2017hydrodynamic}
T. Scaffidi, N. Nandi, B. Schmidt, A. P. Mackenzie and J. E. Moore.
\newblock {\em Phys. Rev. Lett.} {\bfseries 118}, 226601 (2017).

\bibitem{alekseev2016negative}
P. S. Alekseev.
\newblock {\em Phys. Rev. Lett.} {\bfseries 117}, 166601 (2016).

\bibitem{pellegrino2017nonlocal}
F. M. D. Pellegrino, I. Torre and M. Polini.
\newblock {\em Phys. Rev. B} {\bfseries 96}, 195401 (2017).

\bibitem{sherafati2016hall}
M. Sherafati, A. Principi and G. Vignale.
\newblock {\em Phys. Rev. B} {\bfseries 94}, 125427 (2016).

\bibitem{sherafati2019hall}
M. Sherafati and G. Vignale.
\newblock {\em Phys. Rev. B} {\bfseries 100}, 115421 (2019).

\bibitem{berdyugin2019measuring}
A.~I. Berdyugin et al.
\newblock {\em Science} {\bfseries 364}, 162 (2019).

\bibitem{cortijo2015hall}
A. Cortijo, Y. Ferreiro\'s, K. Landsteiner and M. A. H. Vozmediano.
\newblock {\ttfamily arXiv:1506.05136}.

\bibitem{heidari2019hall}
S. Heidari, A. Cortijo and R. Asgari.
\newblock {\em Phys. Rev. B} {\bfseries 100}, 165427 (2019).



\bibitem{read2011hall}
N. Read and E. H. Rezayi.
\newblock {\em Phys. Rev. B} {\bfseries 84}, 085316 (2011).

\bibitem{bradlyn2012kubo}
B. Bradlyn, M. Goldstein and N. Read.
\newblock {\em Phys. Rev. B} {\bfseries 86}, 245309 (2009).

\bibitem{chernodub2017scale}
M. N. Chernodub and M. Zubkov.
{\em Phys. Rev. D} {\bfseries 96}, 056006 (2017).

\bibitem{zhang2020influence}
C. Zhang and M. Zubkov.
{\em Jour. Phys. A: Mathematical and Theoretical} {\bfseries 53}, 195002 (2020).

\bibitem{zubkov2023effect}
M. A. Zubkov and R. A. Abramchuk,
{\em Phys. Rev. D} {\bfseries 107}, 094021 (2023).

\bibitem{zhang2019hall}
C. Zhang and M. Zubkov.
{\em JETP Letters} {\bfseries 110}, 487 (2019).

\bibitem{suleymanov2019wigner}
M. Zubkov.
{\em Nucl. Phys. B} {\bfseries 938}, 171 (2019).

\bibitem{supmat}
\newblock {See Supplementary Material.}





\bibitem{katsnelson2013euler}
M. Katsnelson, G. Volovik, and M. Zubkov.
\newblock{\em Ann. Phys.} {\bfseries 331}, 160 (2013).

\bibitem{leyva2015generalizing}
M. Oliva-Leyva and G. G. Naumis.
\newblock{\em Phys. Lett. A} {\bfseries 379}, 2645 (2015).



\bibitem{volovik2015emergent}
G. E. Volovik and M. A. Zubkov.
\newblock {\em Ann. Phys.} {\bfseries 356}, 255 (2015).

\bibitem{zubkov2015emergent}
M. A. Zubkov and G. E. Volovik.
\newblock {\em J. Phys.: Conf. Ser.} {\bfseries 607}, 012020 (2015).

\bibitem{juan2013gauge}
F. de Juan, J. L. Ma\~nes and M. A. H. Vozmediano.
\newblock {\em Phys. Rev. B} {\bfseries 87}, 165131 (2013).









	


\bibitem{volovik2015scalar}
M. Zubkov.
{\em Phys. Rev. D} {\bfseries 92}, 055004 (2015).

\bibitem{volovik2013nambu}
G. E. Volovik and M. Zubkov.
{\em JETP Letters} {\bfseries 97}, 301 (2013).

\bibitem{vanderbilt2000berry}
D. Vanderbilt.
\newblock {\em J. Phys. Chem. Sol.} {\bfseries 61}, 147 (2000).

\bibitem{resta2007theory}
R. Resta and D. Vanderbilt.
\newblock {``Theory of Polarization: A Modern Approach.''}
in
\newblock {``Physics of Ferroelectrics: A Modern Perspective.''}
Eds. K. Rabe, C. H. Ahn and J.-M. Triscone.
\newblock {\em Springer} (2007).

\bibitem{droth2016piezoelectricity}
M. Droth, G. Burkard and V. M. Pereira.
\newblock {\ttfamily arXiv:1604.01512v2}.

\bibitem{qhemaik2}
M. Selch, M. A. Zubkov, S. Pramanik and M. Lewkowicz.
\newblock {\em Ann. Phys.} {\bfseries 482}, 170202 (2025).

\bibitem{selch2026nonrenormalization}
M. Selch.
\newblock {\em Ann. Phys.} {\bfseries 494}, 170650 (2026).

\bibitem{zubkov2018momentum}
M. Zubkov.
{\em Ann. Phys.} {\bfseries 393}, 264 (2018).

\bibitem{zubkov2012momentum}
M. Zubkov and G. Volovik.
{\em Nucl. Phys. B} {\bfseries 860}, 295 (2012).

\bibitem{volovik2017standard}
G. Volovik and M. Zubkov.
{\em New Jour. Phys.} {\bfseries 19}, 015009 (2017).

\bibitem{zubkov2017topology}
M. Zubkov.
{\em JETP Letters} {\bfseries 106}, 172 (2017).

\bibitem{bakker1999central}
B. Bakker, A. Veselov, and M. Zubkov.
{\em Phys. Lett. B} {\bfseries 471}, 214 (1999).

\bibitem{abramchuk2018anatomy}
R. Abramchuk, Z. Khaidukov, and M. Zubkov.
{\em Phys. Rev. D} {\bfseries 98}, 076013 (2018).

\bibitem{bakker2005standard}
B. Bakker, A. Veselov, and M. Zubkov.
{\em Phys. Lett. B} {\bf 620}, 156 (2005).

\bibitem{Hatsugai}
Y. Hatsugai, T. Fukui, and H. Aoki. 
{\em Phys. Rev. B}
{\bfseries 74}, 205414 (2006).

\bibitem{selch2026valley}
M. Selch.
\newblock {\ttfamily arXiv:2607.08648}.

\bibitem{kim2018accurate}
Kim et al.
\newblock{\em Nano Lett.} {\bfseries 18}, 7732 (2018).

\bibitem{sadeghi2023tunable}
Sadeghi et al.
\newblock{\em Nature} {\bfseries 617}, 282 (2023).

\bibitem{mechelen2019viscous}
T. Van Mechelen and Z. Jacob.
\newblock{ \ttfamily arXiv:1910.14288}.





\bibitem{kim2025viscous}
Kim et. al.
\newblock {\em Phys. Rev. B} {\bfseries 112}, L201404 (2025).

\bibitem{mechelen2021optical}
T. Van Mechelen, W. Sun and Z. Jacob.
\newblock {\em Nat. Com.} {\bfseries 12}, 4729 (2021).













\bibitem{SonWingate2006}
D. T. Son and M. Wingate.
\newblock {\em Ann. Phys.} {\bfseries 321}, 197 (2006).
\newblock {\ttfamily arXiv:cond-mat/0509786}.

\bibitem{qhemaik}
M. Selch, M. Suleymanov, C. Z. Zhang and M. A. Zubkov.
\newblock{\em Phys. Rev. B} {\bfseries 107}, 245105 (2023).

\bibitem{Tong:2016kpv}
D. Tong. 
\newblock {\ttfamily arXiv:1606.06687}.

\end{thebibliography}
\end{document}